%% file: main.tex
\documentclass{article}
\usepackage{siunitx}
\usepackage{booktabs}
\usepackage{subcaption}
\usepackage[most]{tcolorbox}
\usepackage{tikz}
\usetikzlibrary{arrows.meta, positioning}
\usepackage{rotating}
\usepackage{tabularx}
\usepackage{url}
\usepackage{hyperref}
\usepackage{float}
\usepackage{enumitem}

\usepackage[numbers]{natbib}
\usepackage{tfrupee}
\usepackage{authblk}
\newcommand{\keywords}[1]{\par\addvspace\baselineskip
\noindent\textbf{Keywords:}\enspace\ignorespaces#1}
\usepackage{amsthm}
\theoremstyle{remark} 
\newtheorem*{remark}{Remark} 
\usepackage{amssymb}
\usepackage{ragged2e}
\usepackage{geometry}
\title{Event-Time Anchor Selection for Multi-Contract Quoting}

\begin{document}

\author[1,2]{Aditya Nittur Anantha}
\author[1]{Shashi Jain}
\author[3]{Shivam Goyal}
\author[3]{Dhruv Misra}

\affil[1]{Department of Management studies, Indian Institute of Science, Bengaluru, Karnataka, 560012, India}
\affil[2]{SigmaQuant Technologies Pvt. Ltd, Bengaluru, Karnataka, 560066, India}
\affil[3]{AlgoQuant Technologies Ltd., Bengaluru, Karnataka, 560066, India}

\maketitle

\begin{abstract}
When quoting across multiple contracts, the sequence of execution can be a key driver of implementation shortfall relative to the target spread~\cite{bergault2022multi}. We model the short-horizon execution risk from such quoting as variations in transaction prices between the initiation of the first leg and the completion of the position. Our quoting policy anchors the spread by designating one contract ex ante as a \emph{reference contract}. Reducing execution risk requires a predictive criterion for selecting that contract whose price is most stable over the execution interval. This paper develops a diagnostic framework for reference-contract selection that evaluates this stability by contrasting order-flow Hawkes forecasts with a Composite Liquidity Factor (CLF) of instantaneous limit order book (LOB) shape. We illustrate the framework on tick-by-tick data for a pair of NIFTY futures contracts. The results suggest that event-history and LOB-state signals offer complementary views of short-horizon execution risk for reference-contract selection.

\keywords{High Frequency Trading, Execution, Slippage, Market Microstructure, Liquidity, Quoting}
\end{abstract}

\section{Introduction}

Automated trading in electronic limit order books produces a dense, time-stamped order flow that motivates an event-time rather than calendar-time treatment of short-horizon execution risk~\cite{brogaard2014high, aldridge2013high, clapham2022hft}. Advances in networking, matching-engine design, and co-located infrastructure have enabled rapid order placement and adjustment at microsecond horizons.

In liquid venues, this latency profile is now routine. With increased adoption of low-latency infrastructure across venues, the marginal value of further latency reductions has declined. Consequently, execution has shifted from predominantly aggressive liquidity-taking policies to adaptive quoting policies~\cite{hasbrouck2013low}. Quoting is effective because it is anticipatory: placements and updates are conditioned on short-horizon forecasts of the microstructure state, rather than solely on the current tick or current limit order book state. In practice, quoting systems place and update limit orders on one or both sides of the LOB using estimated near-term trajectories of order flow, queue depth, and price continuity \cite{avellaneda2008high, foucault2005limit}. Design goals include lowering inventory variance, capturing bid–ask spread, and maintaining price competitiveness under latency and reliability constraints. Quoting performance, therefore, depends on forecasting of short-horizon order flow and the probability of adverse selection, both of which influence spreads and trading costs \cite{GlostenMilgrom1985, glosten1987components}. Accordingly, in this paper we treat predictive modeling as a central component of quoting and use this perspective to motivate the choice of a reference instrument (basis) in multi-contract settings.

No-arbitrage requirements jointly constrain the prices of related contracts, necessitating coordinated pricing and execution across individual legs \cite{bertsimas2001stochastic}. Accordingly, we adopt multi-contract arbitrage (see Section~\ref {sec:arbitrage_example} for a detailed example) in a diagnostic setting: in liquid markets, arbitrage opportunities are short-lived and spreads are thin. Even small reductions in execution cost translate directly into realized P\&L \cite{kearns2010empirical, hasbrouck2013low}. As illustrative structures, calendar spreads, butterflies, and iron condors are routinely used to express short-horizon views under thin spreads and tight constraints \cite{hull2018options, buehlmaier2019option, macbeth2011trading}. This sensitivity makes reference-leg selection a measurable problem of near-term basis stability, which we evaluate using only information available at the start of each decision interval.

In multi-contract arbitrage, a quoting policy must select a reference leg and price the target leg’s quotes relative to that reference. Operationally, the strategy posts limit orders on the target leg while monitoring a designated reference price. Once the target order fills, the strategy submits a marketable limit order on the reference leg to complete the position.\footnote[1]{On the National Stock Exchange (NSE) of India, market orders are disallowed for algorithmic systems; orders must include explicit price limits, therefore, our implementation uses marketable limit orders on the reference leg.} This design places orders in advance rather than reacting after opportunities arise, aiming to capture a trader-defined spread while controlling execution cost. In practice, quote prices are functions of the observed mid-price of the target leg and the contemporaneous reference price. The approach presumes that the reference remains sufficiently stable over the fill horizon. Instability in the basis results in deviation from target spread and reduced profitability \cite{bergault2022multi}.In this setting, execution risk is amplified because the reference leg establishes the pricing basis while other legs await fills. Our focus is on the information available at decision time: a temporal component based on multivariate Hawkes processes (order-flow clustering and cross-excitation) and a contemporaneous LOB-state component aggregated as a Composite Liquidity Factor (CLF). In this design, the selection of reference-legs is evaluated from two differing information sets.

\begin{remark}
We study reference selection as an \textit{informational} problem for execution–cost control, not as a prescriptive quoting rule. Two information sets are considered at decision time:
(i) \textit{temporal interaction} in order–flow events, modeled with a multivariate Hawkes process (clustering and cross–excitation), and
(ii) \textit{contemporaneous LOB state}, summarized as a Composite Liquidity Factor (CLF) derived from depth and price discontinuities.
Decisions are evaluated in a rolling, walk–forward, out–of–sample design, with each interval’s choice formed from the information available at its beginning. Evaluation is anchored by a hindsight lower bound on deviation from the target spread, which provides a common yardstick for comparing heterogeneous rules.
\end{remark}

\subsection{Literature and existing frameworks}

A large body of literature models optimal quoting under inventory and adverse selection constraints in dealer and electronic markets \cite{HoStoll1981, GlostenMilgrom1985}. In electronic limit order books, Avellaneda and Stoikov \cite{avellaneda2008high} formulate a stochastic control model for dynamic bid/ask placement that trades off expected return against inventory risk, while Cartea \textit{et al.} \cite{cartea2015algorithmic} survey algorithmic trading frameworks that include inventory-aware quoting. Related microstructure work shows how quoting behavior shapes liquidity and price formation \cite{foucault2005limit}. Collectively, these studies support dynamic quotes that respond to inventory and volatility. However, the canonical formulations are predominantly single-asset and do not address the choice of reference instrument in multi-contract quoting.

\subsubsection*{From single-asset quoting to multi-contract reference-leg selection}
Multi-leg spreads (e.g., futures calendar spreads and option combinations) raise execution challenges that single-asset models do not address. Prior work extends optimal control to multi-asset settings, including stochastic programs for arbitrage portfolios \cite{bertsimas2001stochastic} and closed-form policies for multi-asset market making with joint inventory control \cite{bergault2018closed}. These studies show potential efficiency gains from coordinated quoting across related instruments. In practice, legs rarely execute simultaneously: quoting policies place limit orders on the non-reference leg(s) and execute the reference leg last with a marketable limit order once those orders fill \cite{aldridge2013high}. This sequential execution creates a short fill horizon in which adverse price moves against the reference cause the realized spread to deviate from the targeted spread. Multi-asset execution frameworks also address cross-asset coordination \cite{bergault2022multi}. However, while multi-contract quoting has been studied, the specific question of sequencing risk under different choices of the reference leg is, to our knowledge, not established in the literature.

A second line of work studies real-time features derived from LOB state. The configuration of the LOB, including depth at best quotes, LOB imbalance, and the spread, contains predictive information about short-horizon price changes \cite{cont2010price, bouchaud2009markets}. In a Markovian LOB setting, Cont \textit{et al.} show that transition probabilities depend on volumes at the best quotes and related state variables \cite{cont2010price}. In practice, limit order book features are routinely incorporated as decision inputs in high-frequency execution and market-making systems \cite{cartea2015algorithmic, hasbrouck2013low}. Several studies propose composite indicators to summarize liquidity conditions by combining depth, spread, and imbalance. Building on this idea, we construct a Composite Liquidity Factor (CLF) that aggregates instantaneous LOB features into a scalar proxy for near-term price stability derived from depth and price discontinuities, and we use it as a real-time decision input in our framework. To our knowledge, prior studies analyze gap statistics and their relation to jumps and impact, but do not aggregate cross-level gaps with depth into a single factor for reference-leg selection in multi-contract quoting \cite{bouchaud2002statistical, cont2010price,blanchet2015continuous, cartea2015algorithmic}.

A third line of work models market activity in event time rather than clock time. Auto-regressive Conditional Duration (ACD) models show that trade durations exhibit serial dependence that can be exploited for short-horizon forecasting \cite{EngleRussell1998}. Multivariate self-exciting point processes extend this approach to order-flow interactions: Hawkes models capture clustering and cross-excitation between trades and quote updates and have been applied extensively in high-frequency settings \cite{Bowsher2007, bacry2015hawkes, chavez2012hawkes, morariu2021state}. Empirical studies report that incorporating self-excitation improves short-horizon predictions of price changes and liquidity shortfalls \cite{hardiman2013predicting}. A practical consideration is the specification of the excitation kernels with both parametric and non-parametric estimators being used in practice \cite{2011a}. In our framework, the Hawkes arrival ratio provides temporal estimates of near-term order-flow intensity and direction, yielding leg-specific forecasts that inform reference-leg selection. Within Hawkes-based LOB models, (i) queue-reactive specifications endogenize excitation via queue-size–dependent terms \cite{wu2019queuereactive, huang2013queuereactive}, while (ii) LOB–dependent specifications let the intensity vary with observable LOB co-variates \cite{mucciante2024orderbookhawkes, protter2024queuehawkesmarkovian}.

In summary, to our knowledge and within the scope of this review, the problem of \textit{dynamic reference-leg selection} for multi-contract spreads remains underdeveloped. To address this gap, we propose a framework that combines a Hawkes-based estimate of imminent relative price stability with a Composite Liquidity Factor (CLF) summarizing contemporaneous LOB state. We evaluate the resulting choices in a rolling, walk-forward, and out-of-sample design.

\subsection{Contribution of This Paper}

This paper presents a general framework for incorporating order flow and LOB state in quoting policy. Specifically, we compare an arrival ratio based on Hawkes processes with a composite liquidity factor (CLF) to study reference-leg selection in multi-contract quoting. Under this framework, the contributions of this paper are summarized as follows:

\begin{enumerate}
  \item \textbf{Arrival ratio ($\rho$)}: A leg-specific, Hawkes-based estimate of short-horizon order-flow direction and intensity used for reference-leg selection.
  \item \textbf{Composite Liquidity Factor (CLF)}: A scalar summary of limit order book state (depth and cross-level price gaps) used as a proxy for short-term price stability in reference-leg selection.
  \item \textbf{Benchmark oracle}: A deviation from the target spread metric, compared to an infeasible hindsight lower bound (oracle), enabling like-for-like comparison across heterogeneous rules.
\end{enumerate}

\subsection{Execution Structure and Reference Leg Selection Problem}
\label{sec:arbitrage_example}
Among the many available multi-contract strategies that employ quoting, this study uses the calendar-spread strategy. The calendar spread offers two advantages: it has the minimal number of legs for a multi-contract design (two), and the only variation between the two contracts is the date of expiration on the same underlying. This concentrates execution effects in the traded spread with fewer confounding factors. Execution of calendar spreads involves taking offsetting positions in two futures contracts with different expiries. We define the reference-leg selection problem as choosing, at each decision time, that contract whose short-horizon dynamics will anchor the spread quote while the other leg awaits execution.

\subsubsection{Quoting Mechanics and Execution Risk}

In a futures calendar spread, the traded entity is the price difference, or \textit{spread}, between the near-term and far-term contracts. Table~\ref{tab:lob_t0} presents bid and ask prices for NIFTY futures expiring in February 2022 and March 2022. In typical carry regimes, near-term contract prices are below the farther-term contract, consistent with the cost-of-carry term structure. 

To illustrate the mechanics of a calendar spread, we consider execution under a liquidity-taking setup. The trader submits \textit{marketable limit orders} to buy the near-term contract at the best ask and to sell the far-term contract at the best bid, achieving immediate fills. With the sign convention that the calendar spread is quoted as far minus near, this entry corresponds to a long-near/short-far position with a positive spread at execution.\footnote{Throughout, references to liquidity-taking strategies indicate marketable limit orders (not market orders). See Remark~\ref{remark:aggressive_limit_orders}.}

\begin{remark}
\label{remark:aggressive_limit_orders}
\textbf{(On liquidity-taking via marketable limit orders)}\\
In this paper, liquidity-taking execution is implemented exclusively with \textit{marketable limit orders} (also referred to as “aggressive” limit orders), i.e., limit orders priced to execute immediately against standing best quotes. We do not use market orders. All references to liquidity-taking strategies should be interpreted accordingly.
\end{remark}

We adopt the following notational conventions throughout: 
\( p(t) \) denotes the quoted limit order book price of a futures contract as a function of time \( t \), with superscripts \( b \) and \( a \) indicating bid and ask, respectively. Subscripts \( c \) and \( n \) refer to the near (current) and far (next) month contracts. 
\( \tilde{p}(t) \) denotes an executed (traded) price at time \( t \), distinguishing it from quoted prices. 
We define the calendar spread as \( S(t) = p_n(t) - p_c(t) \) (next minus current). 
The realized spreads at entry and exit are denoted \( \tilde{S}_{\text{entry}} \) and \( \tilde{S}_{\text{exit}} \). 
The net realized profit is
\( \Pi = \tilde{S}_{\text{entry}} - \tilde{S}_{\text{exit}} \).
We let \( t_0 \) denote the time of spread entry and \( t_1 \) the time of spread exit.

Suppose the trader anticipates that, over the intended holding interval, the calendar spread will narrow (for example, owing to temporary changes in cost of carry or relative demand). Under the liquidity-taking setup described above and the same sign convention, the trader enters a long-near/short-far position and plans to exit once the realized spread has tightened by the targeted amount. The trader initiates the calendar spread by:
\begin{enumerate}
    \item Buying the near-term (February) futures contract at the executed ask price \( \tilde{p}_c^a(t_0) = 17,458.55 \),
    \item Selling the far-term (March) futures contract at the executed bid price \( \tilde{p}_n^b(t_0) = 17,514.20 \),
\end{enumerate}
at time \( t = t_0 \).

The \textit{entry spread} realized at time \( t_0 \) is computed as:
\begin{equation}
    \tilde{S}_{\text{entry}} = \tilde{p}_n^b(t_0) - \tilde{p}_c^a(t_0),
\end{equation}
\[
\tilde{S}_{\text{entry}} = 17,514.20 - 17,458.55 = 55.65.
\]

If the market evolves as expected and the spread narrows, the trader can exit the spread profitably. Suppose at time \( t = t_1 \), the February contract's bid price rises to \( \tilde{p}_c^b(t_1) = 17,460.00 \) and the March contract's ask price falls to \( \tilde{p}_n^a(t_1) = 17,510.00 \). Then, to close the position, the trader:
\begin{enumerate}
    \item Sells the February contract at \( \tilde{p}_c^b(t_1) = 17,460.00 \),
    \item Buys back the March contract at \( \tilde{p}_n^a(t_1) = 17,510.00 \).
\end{enumerate}

The \textit{exit spread} realized at time \( t_1 \) is:

\begin{equation}
    \tilde{S}_{\text{exit}} = \tilde{p}_n^a(t_1) - \tilde{p}_c^b(t_1),
\end{equation}
\[
\tilde{S}_{\text{exit}} = 17,510.00 - 17,460.00 = 50.00.
\]

The net realized profit per unit traded is given by:

\begin{equation}
    \Pi = \tilde{S}_{\text{entry}} - \tilde{S}_{\text{exit}},
\end{equation}
\[
\Pi = 55.65 - 50.00 = 5.65.
\]

\subsubsection{Execution using quoting}

The same spread trade can be executed through a quoting strategy. In this approach, the trader begins with a predefined target spread \( S \). Rather than accepting the current market spread as fixed, the trader aims to minimize the cost of exit and maximize the profitability of the entry by selectively quoting one leg of the spread.

To execute the strategy, the price of one contract is designated as the \textit{reference price}, and a limit order is placed on the other contract at a price derived as a function of the reference price and the target spread \( S \). Upon the limit order being filled, a corresponding marketable limit order (\emph{crossing the book}) is triggered on the reference contract to complete the spread execution.

In pure liquidity-taking strategies, where immediate aggressive orders are fired on both legs simultaneously, the execution risk is considered negligible. This assertion is valid in markets that support atomic two-leg and three-leg Immediate-Or-Cancel (IOC) order types, which guarantee that either both legs execute simultaneously or neither leg executes. At the National Stock Exchange (NSE) of India, such IOC multi-leg orders are supported. Consequently, when using simultaneous liquidity-taking across legs, the execution of the trader's view on the spread occurs with atomicity, and execution risk can be considered zero.

In quoting strategies, however, execution risk is non-zero. This arises because of the latency between the limit order on the quoting leg being filled and the marketable limit order on the reference leg being fired and executed. During this time window, the reference price may change adversely. 

As a result, the realized spread at time \( t \), denoted \( \tilde{S}(t) \), may differ from the trader’s intended spread \( S \). In certain market conditions such as contango, the farther-month futures price may trade below the nearer-month price, leading to a negative target spread.\footnote{See Remark~\ref{remark:reverse_arbitrage} for a discussion on contango markets and reverse spreads.} The slippage associated with quoting, denoted \( \delta_S(t) \), is defined as:

\begin{equation}
    \delta_S(t) = |S - \tilde{S}(t)|.
\end{equation}

A critical aspect of quoting across contracts is the choice of which contract to reference. The choice directly impacts both the realized cost at entry and the realized profit at exit. The core forecasting problem addressed in this paper is selecting the reference leg that minimizes the probability of unfavorable price changes during the execution window.

\begin{remark}
\label{remark:reverse_arbitrage}
\textbf{(On Reverse Arbitrage and Contango Markets)}\\
In certain market conditions, such as contango, the futures price of a farther-month contract may trade below that of a nearer-month contract. This results in a negative intended spread \( S \) for calendar spread strategies. The slippage definition \( \delta_S(t) = |S - \tilde{S}(t)| \) naturally accommodates both positive and negative target spreads.
\end{remark}

\subsubsection{Execution using limit orders}

Market makers frequently employ quoting strategies that involve placing limit orders on both the bid and ask sides of the limit order book, aiming to profit from short-term fluctuations in the bid-ask spread, especially during phases of spread expansion or contraction. However, in the context of futures rollover trades, simultaneously quoting both legs of a multi-contract strategy increases execution risk. The resulting fill asymmetry can lead to a significant deviation between the realized spread \( \tilde{S}(t) \) and the trader’s intended spread \( S \), reflected in higher slippage \( \delta_S(t) \).

Another critical consideration when designing quoting policies is whether the objective is to minimize execution risk alone, or whether it must also meet minimum trading volume constraints over time. While the framework presented in this paper can be extended to symmetric quoting across multiple contracts, our focus is specifically on \textit{asymmetric multi-contract quoting strategies}, in which limit orders are placed only on a subset of contracts and priced using reference prices derived from a disjoint subset.

\subsection{ Impact of Reference Leg Selection on Execution Slippage}

We consider two futures contracts on the same underlying: the \textit{current-month contract} (denoted by subscript $c$) and the \textit{next-month contract} (subscript $n$). The trader wishes to enter a spread trade at time $t_0$ with a target spread of \( S(t_0) = 58.5 \) INR per contract. We adopt the following time Convention:
\begin{itemize}[label={}]
    \item $t_0$: time of initial quote placement.
    \item $t_1$: time when the quoting leg limit order is executed.
    \item $t_2$: time when the reference leg (aggressive order) is executed.
\end{itemize}
The interval $[t_1, t_2]$ is referred to as the \textit{critical interval}. We reiterate here for clarity, the notation Convention:
\begin{itemize}[label={}]
    \item $p^b_k(t)$: best bid price of contract $k \in \{c, n\}$ at time $t$
    \item $p^a_k(t)$: best ask price of contract $k$ at time $t$
    \item $S^{(r)}(t)$: spread implied by reference leg $r \in \{c, n\}$ at time $t$
    \item $\tilde{S}^{(r)}(t_2)$: realized spread at execution
    \item $\delta_S^{(r)}(t_2) = |\tilde{S}^{(r)}(t_2) - S(t_0)|$: slippage
\end{itemize}
All prices and spreads are expressed in INR per contract.

\begin{remark}
\label{remark:slippage}
\textbf{(On Slippage and Execution Timing)}\\
The time difference between the execution of the quoting leg and the reference leg is called the \textit{critical interval}. During this interval, the top-of-book price on the reference leg may change. This affects the realized spread $\tilde{S}^{(r)}(t_2)$, causing deviation from the trader’s target spread $S(t_0)$. We define the slippage incurred as:
\[
\delta_S^{(r)}(t_2) = \left| \tilde{S}^{(r)}(t_2) - S(t_0) \right|
\]
A positive slippage implies favorable price movement during execution; a negative value implies a cost.
\end{remark}

\vspace{0.3cm}
\subsubsection*{Top of the Book at \( t_0 \)}

\begin{table}[!ht]
\centering
\setlength{\tabcolsep}{4pt}\renewcommand{\arraystretch}{1.05}
\begin{subtable}[t]{0.48\linewidth}
\captionsetup{justification=raggedright,singlelinecheck=false}
\centering
\scriptsize
\caption{NIFTY February 2022 (Prices in INR; sizes in contracts)}
\begin{tabular}{@{}rr|rr@{}}
\toprule
\textbf{Bid} & \textbf{Qty} & \textbf{Ask} & \textbf{Qty}\\
\midrule
17455   &  50 & 17458.55 &  50\\
17450   &  50 & 17459.65 &  50\\
17401.1 & 100 & 17459.95 & 100\\
17376   & 800 & 17460    & 550\\
17355   & 100 & 17475    & 250\\
\bottomrule
\end{tabular}
\end{subtable}\hfill
\begin{subtable}[t]{0.48\linewidth}
\captionsetup{justification=raggedright,singlelinecheck=false}
\centering
\scriptsize
\caption{NIFTY March 2022 (Prices in INR; sizes in contracts)}
\begin{tabular}{@{}rr|rr@{}}
\toprule
\textbf{Bid} & \textbf{Qty} & \textbf{Ask} & \textbf{Qty}\\
\midrule
17514.2 &  50 & 17516.5 & 150\\
17510.2 & 150 & 17516.55&  50\\
17510.1 & 100 & 17516.7 &  50\\
17509.1 & 150 & 17517.2 &  50\\
17503.6 &  50 & 17523.4 & 200\\
\bottomrule
\end{tabular}
\end{subtable}
\caption{Top 5 LOB levels at \(t_0\) for NIFTY futures}
\label{tab:lob_t0}
\end{table}

From the limit order books in table~\ref{tab:lob_t0}, at time $t_0$, the two implied spreads are:

\begin{equation}
\begin{aligned}
S^{(c)}(t_0) &= p^b_n(t_0) - p^b_c(t_0) \\
             &= 17514.2 - 17455  \\
             &= 59.2
\end{aligned}
\end{equation}
\begin{equation}
\begin{aligned}
S^{(n)}(t_0) &= p^a_n(t_0) - p^a_c(t_0) \\
             &= 17516.5 - 17458.55 \\
             &= 57.95
\end{aligned}
\end{equation}

\vspace{0.3cm}
\subsubsection{Case 1: Current Month as Reference Leg:}
\paragraph{}
In this setup, the trader designates the \textit{current-month contract} as the reference leg. A buy limit order is placed on the \textit{next-month contract} at time \( t_0 \), quoting the best available ask price of \( p^a_n(t_0) = 17514.2 \). This order is executed at time \( t_1 \).

Following this execution, at time \( t_2 \), the trader fires a \textit{marketable limit order} to sell the current-month contract. During the \textit{critical interval} \( [t_1, t_2] \), the top bid on the current month leg improves from 17455 to 17458. Consequently, the aggressive order is executed at the updated price \( p^b_c(t_2) = 17458 \).

The spread realized under this execution path, denoted \( \tilde{S}^{(c)}(t_2) \), is calculated as the difference between the bid on the next month leg (observed at \( t_1 \)) and the updated bid on the reference leg (current month) at \( t_2 \). This spread is \rupee 56.2, leading to a slippage of \rupee 2.3 compared to the target spread of \rupee 58.5.

Despite initially observing a wider possible spread, the trader incurs higher slippage due to adverse price movement during the critical interval.

\begin{table}[!ht]
\centering
\tiny
\setlength{\tabcolsep}{4pt}\renewcommand{\arraystretch}{1.05}

\begin{subtable}[t]{0.48\linewidth}
\captionsetup{justification=raggedright,singlelinecheck=false}
\centering
\scriptsize
\caption{NIFTY March 2022 ($t_1$)}\label{tab:near_leg_ref_mar_t1}
\begin{tabular}{@{}ll|ll@{}}
\toprule
\textbf{Bid} & Qty & \textbf{Ask} & Qty\\
\midrule
17514.2 & 50  & \textcolor{blue}{17514.2} & \textcolor{blue}{50}\\
17510.2 & 50  & 17516.5                    & 150\\
17510.1 & 150 & 17516.55                   & 50\\
\bottomrule
\end{tabular}
\end{subtable}\hfill
\begin{subtable}[t]{0.48\linewidth}
\captionsetup{justification=raggedright,singlelinecheck=false}
\centering
\scriptsize
\caption{NIFTY February 2022 ($t_2$)}\label{tab:near_leg_ref_feb_t2}
\begin{tabular}{@{}ll|ll@{}}
\toprule
\textbf{Bid} & Qty & \textbf{Ask} & Qty\\
\midrule
\textcolor{red}{17458} & \textcolor{red}{50} & 17458.55 & 50\\
17455                  & 50                  & 17459.65 & 50\\
17450                  & 50                  & 17459.95 & 100\\
\bottomrule
\end{tabular}
\end{subtable}

\caption{LOB snapshots when current month is the reference leg}\label{tab:near_leg_ref}
\end{table}

\begin{equation}
\begin{aligned}
\tilde{S}^{(c)}(t_2) &= p^b_n(t_1) - p^b_c(t_2) \\ 
                     &= 17514.2 - 17458 \\
                     &= 56.2 
\end{aligned}
\end{equation}
\begin{equation}
\begin{aligned}
\delta_S^{(c)}(t_2) &= |56.2 - 58.5| \\
                    &= 2.3 \text{ INR per contract}
\end{aligned}
\end{equation}

\vspace{0.3cm}
\subsubsection*{Case 2: Next Month as Reference Leg:}
\paragraph{}
In this alternative configuration, the trader chooses the \textit{next-month contract} as the reference leg. A buy limit order is placed on the \textit{current-month contract} at time \( t_0 \), quoting the best available ask price of \( p^a_c(t_0) = 17458.55 \). This order is filled at time \( t_1 \).

At \( t_2 \), the trader fires a marketable limit order on the next-month contract to sell at market. During the critical interval, the top ask price on the next-month contract shifts slightly from 17516.5 to 17516.55, reflecting minor upward movement. The order is thus executed at the updated price \( p^a_n(t_2) = 17516.55 \).

The realized spread \( \tilde{S}^{(n)}(t_2) \), computed as the difference between the ask on the reference leg (next month) at \( t_2 \) and the ask on the quoting leg (current month) at \( t_1 \), equals \rupee 58.0. This results in a slippage of only \rupee 0.5 from the target spread.

Although the next month leg had a slightly narrower spread at \( t_0 \), the higher short-term stability during the critical interval led to significantly lower execution risk.

\begin{table}[!ht]
\centering
\tiny
\setlength{\tabcolsep}{4pt}\renewcommand{\arraystretch}{1.05}

\begin{subtable}[t]{0.48\linewidth}
\captionsetup{justification=raggedright,singlelinecheck=false}
\centering
\scriptsize
\caption{NIFTY February 2022 ($t_1$)}\label{tab:far_leg_ref_feb_t1}
\begin{tabular}{@{}ll|ll@{}}
\toprule
\textbf{Bid} & Qty & \textbf{Ask} & Qty\\
\midrule
\textcolor{blue}{17458.55} & \textcolor{blue}{50} & 17458.55 & 50\\
17455 & 50 & 17459.65 & 50\\
17450 & 50 & 17459.95 & 100\\
\bottomrule
\end{tabular}
\end{subtable}\hfill
\begin{subtable}[t]{0.48\linewidth}
\captionsetup{justification=raggedright,singlelinecheck=false}
\centering
\scriptsize
\caption{NIFTY March 2022 ($t_2$)}\label{tab:far_leg_ref_mar_t2}
\begin{tabular}{@{}ll|ll@{}}
\toprule
\textbf{Bid} & Qty & \textbf{Ask} & Qty\\
\midrule
17514.2 & 50  & \textcolor{red}{17516.55} & \textcolor{red}{50}\\
17510.2 & 150 & 17516.7                     & 50\\
17510.1 & 100 & 17517.2                     & 50\\
\bottomrule
\end{tabular}
\end{subtable}

\caption{LOB snapshots when next month is the reference leg}\label{tab:far_leg_ref}
\end{table}

\begin{equation}
\begin{aligned}
\tilde{S}^{(n)}(t_2) &= p^a_n(t_2) - p^a_c(t_1) \\
                     &= 17516.55 - 17458.55 \\
                     &= 58.0
\end{aligned}
\end{equation}
\begin{equation}
\begin{aligned}
\delta_S^{(n)}(t_2) &= |58.0 - 58.5| \\
                    &= 0.5 \text{ INR per contract}
\end{aligned}
\end{equation}

\vspace{0.3cm}

\subsubsection*{Summary and Interpretation}

\begin{table}[!ht]
\centering
\setlength{\tabcolsep}{4pt}\renewcommand{\arraystretch}{1.05}

\centering
\captionsetup{type=table,justification=raggedright,singlelinecheck=false}
\begin{tabular}{@{}lcccc@{}}
\toprule
\textbf{Ref. Leg} & \textbf{Ref Price (\(t_0\))} & \textbf{Ref Exec (\(t_2\))} & \textbf{Realized Spread} & \textbf{Slippage (INR)} \\
\midrule
Current Month ($r = c$) & 17455   & \textcolor{red}{17458} & 56.2 & 2.3 \\
Next Month ($r = n$)    & 17516.5 & \textcolor{red}{17516.55} & 58.0 & 0.5 \\
\bottomrule
\end{tabular}
\caption{Reference leg outcomes: reference price at $t_0$ and execution at $t_2$ with realized spread and slippage (in INR) for the current month $(r = c)$ and next month $(r = n).$} 
\label{tab:summary_table}
\end{table}

Although the initial spread was wider when the current month was selected as the reference leg, the final realized spread was lower due to price movement during the critical interval, resulting in higher slippage. Conversely, the next month leg provided better price stability, despite a narrower initial spread, and resulted in lower slippage. 

This highlights the core insight: short-term stability of the reference leg is more important than instantaneous spread width when minimizing execution risk. In the sections that follow, we derive leg-specific stability estimates from LOB state (CLF) and from event history via a Hawkes arrival ratio, and use them to evaluate reference-leg choices.

\section{Problem Formulation: Optimal Reference Leg Selection} \label{sec:problem_formulation}

The preceding empirical comparison illustrated that the choice of reference leg in a calendar spread quoting strategy materially affects the realized spread and slippage. Specifically, even when the initial observed spread favors one contract, short-term price stability during the critical interval can dominate realized execution outcomes. Motivated by these observations, we now formalize the reference leg selection problem as a dynamic decision-making process. At each quoting decision point, the trader must select the contract to treat as the reference leg based on observable market conditions, to minimize expected slippage relative to a target spread constraint. Unlike classical optimization problems, this selection is guided by statistical methods and empirical characteristics of the limit order book rather than by solving an explicit minimization problem. This formulation captures the execution asymmetry inherent in quoting strategies and provides the foundation for systematic reference leg selection based on real-time market data.

Let \( \mathcal{W} = \{w^{(1)}, w^{(2)}, \dots\} \) be a collection of non-overlapping evaluation intervals of the form \( w = [w_1, w_2) \), where \( w_1 \) and \( w_2 \) denote the start and end of each interval respectively.

Iterating over each interval in \( \mathcal{W} \), we consider two classes of filters that guide quoting decisions at time \( w_1 \):
\begin{enumerate}
  \item Filters \( \mathcal{G}^{\mathcal{T}}(w) \) that act on the stream of trades and quotes observed strictly before the start of the interval.
  \item Filters \( \mathcal{G}^{\mathcal{B}}(w) \) that operate on the LOB snapshot observed at the start of the interval.
\end{enumerate}

Each filter selects or scores relevant inputs from its respective domain, producing a filtered representation that will be used to assess the quoting viability of each contract leg.

Each filter is paired with a quoting function that maps the filtered representation to a real-valued score for each candidate leg. The tick-based quoting function \( \mathcal{Q}^{\mathcal{T}} \) operates on the output of \( \mathcal{G}^{\mathcal{T}}(w) \), while the LOB-based quoting function \( \mathcal{Q}^{\mathcal{B}} \) operates on the output of \( \mathcal{G}^{\mathcal{B}}(w) \).

Each quoting function returns a real-valued estimate of expected slippage for each leg \( r \in \{c, n\} \), and selects the leg with the smaller value:
\begin{equation}
\mathcal{Q}^{\mathcal{T}}(w) := \arg\min_{r \in \{c,n\}} \mathbb{E}[\delta_S^{(r)}(w_2) \mid \mathcal{G}^{\mathcal{T}}(w)] \quad \forall w \in \mathcal{W}
\end{equation}

\begin{equation}
\mathcal{Q}^{\mathcal{B}}(w) := \arg\min_{r \in \{c,n\}} \mathbb{E}[\delta_S^{(r)}(w_2) \mid \mathcal{G}^{\mathcal{B}}(w)] \quad \forall w \in \mathcal{W}
\end{equation}

Here, \( \delta_S^{(r)}(w_2) \) denotes the realized slippage incurred on leg \( r \) at the end of the interval, and the filters \( \mathcal{G}^{\mathcal{T}}(w) \) and \( \mathcal{G}^{\mathcal{B}}(w) \) encode the information available at decision time.

The quoting decisions are finally expressed as binary indicators:
\begin{equation}
\chi^{\mathcal{T}}(w) := \mathbb{I} \left\{ \mathcal{Q}^{\mathcal{T}}(w) = n \right\}, \quad
\chi^{\mathcal{B}}(w) := \mathbb{I} \left\{ \mathcal{Q}^{\mathcal{B}}(w) = n \right\}
\end{equation}

where a value of 1 indicates that the quoting rule recommends the next-month leg, and 0 indicates the current-month leg.

We denote by \( \chi^m(w) \in \{0,1\} \) the market-optimal reference leg decision for interval \( w \), defined as the choice that yields the lowest realized slippage in hindsight. This oracle benchmark provides a reference against which to evaluate the quality of decisions made by each quoting rule.

\begin{remark}
    We report realized spread slippage because the remaining cost components (exchange fees, rebates, and longer–horizon opportunity costs) are invariant to the reference–leg choice under our setup: a single venue and identical contract multipliers. These terms therefore add an interval–level constant and do not affect the ranking of \( \chi^{\mathcal{B}} \) versus \( \chi^{\mathcal{H}} \). In market settings with asymmetric fee schedules, cross–venue routing, or role–dependent fills, these terms need not be invariant. Integrating them is orthogonal to our information comparison and is left for future work.
\end{remark}

\begin{tcolorbox}[colframe=black, colback=white!97!gray, boxrule=0.4pt, arc=2pt, left=4pt, right=4pt, top=4pt, bottom=4pt, width=\columnwidth]
\textbf{Problem Definition: Evaluation of Reference Leg Selection Rules.}

Let \( \mathcal{W}=\{w\} \) be non--overlapping evaluation intervals \( w=[w_1,w_2) \).
For each \( w \), let \( \chi^{\mathcal{B}}(w), \chi^{\mathcal{H}}(w) \in \{0,1\} \) denote the binary decisions produced at \( w_1 \) by the CLF--based rule and the Hawkes--based rule, respectively (value \(1\) selects the next--month leg, value \(0\) selects the current--month leg).
Let \( \delta_S^{(r)}(w_2) \) be realized slippage on leg \( r\in\{c,n\} \) over \( [w_1,w_2) \).
Given the market--optimal oracle decision
\[
\chi^{m}(w) \;:=\;
\begin{cases}
1, & \text{if } \delta_S^{(n)}(w_2) < \delta_S^{(c)}(w_2),\\
0, & \text{if } \delta_S^{(c)}(w_2) \le \delta_S^{(n)}(w_2),
\end{cases}
\]
which breaks ties in favor of \(c\) deterministically, we define the agreement score:

\[
\mathcal{A}(\chi, \chi^m) := \frac{1}{|\mathcal{W}|} \sum_{w \in \mathcal{W}} \mathbb{I} \left\{ \chi(w) = \chi^m(w) \right\}
\]

Compute and compare the agreement scores \( \mathcal{A}(\chi^{\mathcal{T}}, \chi^m) \) and \( \mathcal{A}(\chi^{\mathcal{B}}, \chi^m) \).\\

\textit{Note: Decisions \( \chi^{\mathcal{B}}(w) \) and \( \chi^{\mathcal{H}}(w) \) are formed using only information timestamped strictly before \( w_1 \). The realized slippage on \( [w_1,w_2) \) is used only for evaluation and to define \( \chi^{m}(w) \).}
\end{tcolorbox}

\section{Methodology} \label{sec:methodology}

The decision rules introduced in Section~\ref{sec:problem_formulation} are evaluated against an oracle benchmark. In this paper, the oracle is an infeasible hindsight lower bound on deviation from the target spread: for each interval \( w \), the oracle decision \( \chi^{m}(w) \) selects the leg (current or next month) that would have minimized realized slippage on \([w_1,w_2)\). We use this hindsight choice as an evaluation label for scoring data-driven quoting rules formed from observable signals. The methodology implements \( \chi^{\mathcal{T}}(w) \) and \( \chi^{\mathcal{B}}(w) \), instantiates \( \chi^{m}(w) \) from realized outcomes, and computes the agreement score defined in the Problem Definition of Section~\ref{sec:problem_formulation} as the primary evaluation metric. All decisions use only information timestamped strictly before \( w_1 \); outcomes on \([w_1,w_2)\) are used solely for evaluation.

\subsection{Execution Design and Quoting Framework}
This subsection specifies how the executable quoting price is computed \emph{after} the reference leg is chosen: we restate the notation for completeness, benchmark against the no-arbitrage spread, and then express the executable price as a VWAP on the selected reference leg. We frame the problem as a calendar-spread rollover. A common configuration is \emph{buying the spread}: selling the current-month contract and buying the next-month contract simultaneously. Execution requires a side-conditional decision—whether the bid in the current-month or the ask in the next-month offers better instantaneous liquidity for the required size. We treat this as a reference-leg selection problem and evaluate the choice using deviation from the target spread as per the oracle benchmark (Section~\ref{sec:clf_forecasting}).

Let \( F_c \) denote the futures contract expiring in the current-month (priced \( p_c(t) \) at time \( t \)) and \( F_n \) the next-month contract (priced \( p_n(t) \)). Let \( q \) be the quantity to trade and let \( S \) denote a fixed target-spread threshold (a maximum acceptable cost per unit, distinct from the time-varying spread \( S(t)=p_n(t)-p_c(t) \)). A calendar spread trade sells \( q \) units of \( F_c \) and buys \( q \) units of \( F_n \) subject to the constraint \( p_n(t) - p_c(t) \le S \). We next benchmark this threshold against the theoretical spread implied by no-arbitrage carry. Conditional on the reference-leg decision, the executable quoting price is taken as the volume-weighted average of the quoted prices required to fill \(q\) on the chosen leg at time \(t\). The next paragraph makes this VWAP construction explicit.

In an efficient, arbitrage-free market with risk-free rate \(R\) and identical multipliers on the current- and next-month contracts (ignoring dividends/funding spreads), the cost-of-carry relation implies
\( p_n(t) = e^{R (t_n - t_c)}\, p_c(t) \), where \(t_c\) and \(t_n\) denote the expiry times of the current- and next-month contracts, respectively, and \(t_n - t_c\) is measured in years consistent with the compounding convention of \(R\). This yields the model-implied spread
\[ 
    S_m(t) = p_c(t)\,\big(e^{R (t_n - t_c)} - 1\big) 
\].
which may deviate in practice due to mispricing or market frictions. These frictions motivate quoting policies that dynamically adjust order prices to maintain execution costs below \( S \).

We consider two quoting configurations:

\paragraph{Case 1: Using \( F_c \) as reference.}
Conditional on the reference-leg choice, we now compute the executable quoting price on the chosen leg using a VWAP of the LOB levels needed to fill \( q \) at time \( t_0 \).

The price \( p_c^r(t) \) is computed as the VWAP over the top \( \nu \) levels on the bid side of \( F_c \)'s LOB:
\[
p_c^r(t) = \frac{\sum_{i=1}^\nu p_c^{b,i} q_c^{b,i}}{\sum_{i=1}^\nu q_c^{b,i}},
\]
with quoting price on \( F_n \) set to:
\[
p_n^l(t) = S - p_c^r(t).
\]

\paragraph{Case 2: Using \( F_n \) as reference.}
Conditional on selecting the next-month contract as the reference leg, the reference leg is executed via a marketable limit order at time \( t \). The executable reference price \( p_n^r(t) \) is taken as the VWAP of the ask-side levels of the limit order book of \( F_n \) required to fill the target size \( q \):
\[
p_n^r(t) = \frac{\sum_{i=1}^\nu p_n^{a,i} \, q_n^{a,i}}{\sum_{i=1}^\nu q_n^{a,i}},
\]
where \( \nu \) is the smallest depth index such that \( \sum_{i=1}^\nu q_n^{a,i} \ge q \); the last level is partially consumed if needed. The linked quoting price on the current-month contract is then set to
\[
p_c^l(t) = p_n^r(t) - S,
\]
so that the quoted spread equals the target \( S \) under the convention \( S(t) = p_n(t) - p_c(t) \). Here, the superscripts \( r \) and \( l \) denote the reference-leg VWAP and the linked quote, respectively. (If the trading direction requires selling \( F_n \), the construction is symmetric with bid-side levels.)

The realized spread \( \tilde{S}(t) \) is computed from the resulting trade prices after quoting and executing the reference leg with a marketable limit order. Consequently, the choice of reference leg affects execution cost and slippage, and will be evaluated against the oracle benchmark in the next subsection.

\subsection{Benchmarking Framework for quoting policies}
\label{sec:market_benchmark}

To compare quoting rules, we define a market-derived oracle decision \( \chi^m(w) \in \{0,1\} \) based on actual executed spread trades.

Let \( \mathcal{E} = \{T, C, \text{PDM}\} \) denote the set of event types observed in tick data, where \( T \) represents trades, \( C \) represents cancellations, and \( \text{PDM} \) denotes price downward modifications. 

The tick-based filter \( \mathcal{G}^{\mathcal{T}}(w) \) returns the set of valid passive-aggressive trade pairs identified within the interval \( w = [w_1, w_2) \), extracted from tick history and used to construct the benchmark reference leg decision.

\begin{figure}[h]
\centering
\begin{tikzpicture}[scale=0.8]

  \draw[->] (0,0) -- (8.2,0) node[anchor=west] {Time};

  \draw[dashed] (2,0) -- (2,3.5);
  \draw[dashed] (6.7,0) -- (6.7,3.5);
  \node at (2,-0.3) {\small $w_1$};
  \node at (6.7,-0.3) {\small $w_2$};
  \node at (5,-1.0) {\textbf{Evaluation window } $w = [w_1, w_2)$};

  \filldraw[red] (2.5,1) circle (2pt);
  \filldraw[red] (4.5,1) circle (2pt);
  \filldraw[red] (6.5,1) circle (2pt);

  \filldraw[blue] (3.0,2.5) circle (2pt);
  \filldraw[blue] (5.5,2.5) circle (2pt);
  \filldraw[blue] (6.0,2.5) circle (2pt);

  \node[red] at (0.3,1) {\small Trades on $F_c$};
  \node[blue] at (0.3,2.5) {\small Trades on $F_n$};

  \draw[->, thick, gray!60!black] (2.5,1) -- (3.0,2.5);
  \draw[->, thick, gray!60!black] (4.5,1) -- (5.5,2.5);
  \draw[<-, thick, gray!60!black] (6.5,1) -- (6.0,2.5);

  \node[align=left] at (5,3.2) {\small Passive-aggressive \\
                                  \small trade pairs};

  \node at (10.1,1) {}; 

\end{tikzpicture}
\caption{Illustration of oracle construction using passive-aggressive trade pairs within interval \( w = [w_1, w_2) \). The first-leg execution determines the oracle decision \( \chi^m(w) \).}
\label{fig:benchmark_diagram}
\end{figure}

\begin{remark}
\textbf{(Trade-Only Basis for oracle Construction)} \\
Only Trade events are used to construct the oracle decision \( \chi^m(w) \). Events of type Cancellation or Price Downward Modification are excluded from the filtered set \( \mathcal{G}^{\mathcal{T}}(w) \), ensuring that benchmarking reflects executed spread trades alone.
\end{remark}

For each pair \( (t_1, t_2) \in \mathcal{G}^{\mathcal{T}}(w) \), the realized spread is:
\[
\tilde{S}_m(t) = p_n^{b,T}(t_2) - p_c^{a,T}(t_1)
\]
, where $T$ represents event-type Trade.
The oracle decision is:
\[
\chi^m(w) :=
\begin{cases}
1 & \text{if } F_c \text{ was traded first (ref leg is } F_n) \\
0 & \text{if } F_n \text{ was traded first (ref leg is } F_c)
\end{cases}
\]

Each quoting rule \( \chi(w) \in \{\chi^{\mathcal{T}}(w), \chi^{\mathcal{B}}(w)\} \) is compared to \( \chi^m(w) \) using the agreement score:
\[
\mathcal{A}(\chi, \chi^m) := \frac{1}{|\mathcal{W}|} \sum_{w \in \mathcal{W}} \mathbb{I} \left\{ \chi(w) = \chi^m(w) \right\}
\]

\begin{remark}
\textbf{(On Quantity Normalization)} \\
All observed calendar spread trades in \( \mathcal{G}^{\mathcal{T}}(w) \) are evaluated using the minimum tradable quantity permitted by the venue. This ensures that the realized spread \( \tilde{S}_m(t) \) reflects slippage on a per-unit basis, independent of execution size.
\end{remark}

\paragraph{}%
We define the oracle label \( \chi^m(w) \) by sequentially evaluating tick data for calendar-spread trade pairs, i.e., opposite-direction trades across \(F_c\) and \(F_n\) within each considered window. For each valid pair, we compute the realized spread using trade-derived prices. If multiple valid pairs occur within \(w\), we retain the pair that attains the smallest spread. This yields a hindsight lower bound for evaluation. The leg that trades first in the retained pair determines the oracle decision. The complete filtered set is denoted \( \mathcal{G}^{\mathcal{T}}(w) \), and all benchmarking is grounded in this empirically observed structure.

We now formalize two reference leg selection rules that instantiate the quoting framework introduced in Section \ref{sec:problem_formulation}.

\subsection{Tick-based quoting rule.}
For each interval \( w \in \mathcal{W} \), we define a tick-based quoting rule that estimates the expected slippage of each leg using filtered trade and quote activity observed strictly prior to the quoting time \( w_1 \):
\begin{equation}
\mathcal{Q}^{\mathcal{T}}(w) := \arg\min_{r \in \{c,n\}} \mathbb{E}\left[\delta_S^{(r)}(w_2) \mid \mathcal{G}^{\mathcal{T}}(w)\right]
\end{equation}
The quoting decision is then encoded as a binary rule:
\begin{equation}
\chi^{\mathcal{T}}(w) := \mathbb{I}\left\{ \mathcal{Q}^{\mathcal{T}}(w) = n \right\}
\end{equation}

\subsection{Forecasting Execution Risk with Multivariate Hawkes Processes}
\label{sec:hawkes_forecasting}

Building on the event taxonomy above, we introduce a predictive framework that forecasts short-horizon execution risk across candidate reference legs. We estimate forward-looking arrival ratios with multivariate Hawkes processes and map them to a reference-leg decision. This model-based rule complements the reactive rules presented earlier and serves as a forecasting proxy for the oracle decision \( \chi^m(w) \) defined in the Problem Definition (Section~\ref{sec:problem_formulation}). The next subsection specifies the event sets, sides, and windows used to form these arrival-ratio forecasts.

\subsubsection{Model Structure}

Let \( \mathcal{E} = \{T, C, PDM\} \) denote the set of reference-impacting event types: $\mathcal{T}$ for trades, $\mathcal{C}$ for cancellations, and $\mathcal{PDM}$ for downward price modifications. Let \( Y = \{b, a\} \) represent the bid and ask sides of the limit order book, and \( X \in \{c, n\} \) index the current and next-month contracts. Define the joint index set \( \mathcal{I} = \mathcal{E} \times Y \).

Our goal is to forecast, over a forward interval \( (\tau, \tau + \xi] \), the number of reference-impacting events \( \tilde{N}^{X,Y,r}_{(\tau, \tau+\xi]} \) and the total number of events \( \tilde{N}^{X,Y}_{(\tau, \tau+\xi]} \). From these, we compute the arrival ratio:
\begin{equation}
    \rho^{X,Y}_{(\tau, \tau + \xi]} = \frac{\tilde{N}^{X,Y,r}_{(\tau, \tau + \xi]}}{\tilde{N}^{X,Y}_{(\tau, \tau + \xi]}}
\end{equation}

Given that spread execution involves selling \( F_c \) and buying \( F_n \), the ask side of \( F_c \) and the bid side of \( F_n \) are relevant. We therefore define the reference leg decision based on the forecasted arrival ratios:
\begin{equation}
    \tilde{\chi}_{(\tau, \tau + \xi]} = \mathbb{I} \left\{ \rho^{c,a}_{(\tau, \tau + \xi]} \geq \rho^{n,b}_{(\tau, \tau + \xi]} \right\}
\end{equation}

To maintain consistency with the decision framework introduced in Section~\ref{sec:problem_formulation}, we denote this forecasting-driven rule as \( \mathcal{Q}^{\mathcal{H}} \), with its output defined as:
\[
\chi^{\mathcal{H}}(w) := \tilde{\chi}_{(\tau, \tau + \xi]} = \mathbb{I} \left\{ \rho^{c,a}_{(w)} \geq \rho^{n,b}_{(w)} \right\}, 
\]
\[
\quad \text{where } w = (\tau, \tau + \xi]
\]
Here, \( \mathcal{Q}^{\mathcal{H}} \) maps estimated arrival ratios to a binary reference leg decision over window \( w \in \mathcal{W} \). 
\begin{tcolorbox}[colframe=black!60, colback=white!97!gray, boxrule=0.4pt, arc=2pt, left=4pt, right=4pt, top=4pt, bottom=4pt, width=\columnwidth, enhanced, sharp corners=south]
\textbf{Definition: Hawkes-Derived Feature Mapping.}

\smallskip
\noindent
For each interval \( w = [w_1, w_2) \in \mathcal{W} \), let \( \mathcal{G}^{\mathcal{H}}(w) \in \mathbb{R}^2 \) denote the filtered representation obtained by simulating a fitted Hawkes model on recent tick data to estimate directional arrival ratios. Define:
\begin{align*}
&\tilde{N}^{X,Y}_{(w)} \text{, Simulated total Hawkes event counts on } w \\
&\tilde{N}^{X,Y,r}_{(w)} \text{, Simulated events attributed to leg } \\ &r \in \{c, n\} \\
&\rho^{X,Y}_{(w)} := \frac{\tilde{N}^{X,Y,r}_{(w)}}{\tilde{N}^{X,Y}_{(w)}} \\
&\mathcal{G}^{\mathcal{H}}(w) := \left( \rho^{c,a}_{(w)},\; \rho^{n,b}_{(w)} \right) \\
&\mathcal{Q}^{\mathcal{H}}(w) := \arg\min_{r \in \{c,n\}} \rho^{r,s}(w) \\
&\chi^{\mathcal{H}}(w) := \mathbb{I} \left\{ \mathcal{Q}^{\mathcal{H}}(w) = n \right\}
\end{align*}
\noindent
where \( \rho^{r,s}(w) \) represents the simulated relative arrival intensity for quoting side \( s \) of leg \( r \), and \( n \) refers to the next-month contract.
\end{tcolorbox}

The above formalization defines the Hawkes-based decision rule \( \chi^{\mathcal{H}}(w) \) as a function of simulated forward event counts derived from fitted multivariate Hawkes models. By comparing the relative frequency of reference-price-impacting events across LOB sides, this approach translates microstructural event flow forecasts into a structured quoting decision. This provides a direct, data-driven proxy for short-term execution stability under each reference leg and can be evaluated against the market-optimal decision \( \chi^m(w) \).

\begin{tcolorbox}[colback=gray!3, colframe=black!30, title=Modeling Remark]
This decision rule \( \chi^{\mathcal{H}}(w) \) serves as a forecasting-based analog of the oracle decision \( \chi^m(w) \). It selects the contract whose LOB side is expected to exhibit fewer reference-price-impacting events, thus offering more stable execution.
\end{tcolorbox}

\subsubsection{Multivariate Hawkes Process for Event Forecasting}

We model reference-impacting events using multivariate Hawkes processes \cite{bacry2015hawkes}. For each contract \( X \) and side \( Y \), we define a \( D \)-dimensional Hawkes process indexed by \( i \in \mathcal{I} \). The intensity \( \lambda^i_t \) evolves as:
\begin{equation}
    \lambda^i_t = \mu^i + \sum_{j=1}^{D} \int_0^t \phi_{ij}(t - s) \, dN^j_s
\end{equation}
where \( \phi_{ij}(t) \) captures the excitation effect of events of type \( j \) on type \( i \).

Following \cite{anantha2025}, we estimate separate Hawkes models for:
\begin{itemize}[label={}]
    \item Reference-impacting events only \( N^{X,Y,r} \)
    \item All events \( N^{X,Y} \)
\end{itemize}
Each model is estimated on a rolling basis over historical windows \( (\tau - h, \tau] \), yielding fitted kernels \( \hat{\Phi}^{X,Y,r} \) and \( \hat{\Phi}^{X,Y} \).

\subsubsection{Estimation and Simulation Procedure}
The accuracy of short-term event forecasts under the Hawkes framework depends critically on the choice of kernel function \( \phi_{ij}(t) \), which encodes the temporal excitation effect of past events. We evaluate both parametric and non-parametric approaches to kernel estimation, balancing model interpretability, estimation stability, and flexibility of fit.

Parametric kernels include the exponential and sum-of-exponentials forms, whose functional structures are shown in Table~\ref{tab:parametric_kernels}. These allow for low-dimensional maximum likelihood estimation and are widely used in market microstructure applications due to their analytical tractability.

Non-parametric kernels, by contrast, do not assume a fixed functional form. Instead, we consider two classes of non-parametric estimators: (i) the \textit{expectation-maximization} (EM) method, which iteratively estimates a discretised kernel by maximizing a latent-variable likelihood, and (ii) the \textit{conditional law approach}, which reconstructs the kernel by inverting the conditional intensity relation from observed event times. These methods provide flexible approximations of \( \phi_{ij}(t) \), allowing for data-driven recovery of potentially complex excitation structures.

\begin{table}[!ht]
\centering
\caption{Evaluated Parametric Kernels}
\label{tab:parametric_kernels}
\begin{tabular}{lll}
\toprule
Kernel Type & Functional Form & Method \\
\midrule
Exponential & \( \alpha e^{-\beta t} \) & TICK MLE \\
Sum of exponentials & \( \sum_u \alpha_u e^{-\beta_u t} \) & TICK MLE \\
\bottomrule
\end{tabular}
\end{table}

The estimation and simulation procedures employed here closely follow those developed in our earlier work \cite{anantha2025}. The present formulation adapts those methods to forecast arrival ratios for reference-impacting events, enabling their integration into a structured decision rule.

Hawkes kernels are estimated via maximum likelihood using the method presented in \cite{ozaki1979mle}. For parametric kernels, we use exponential and sum-of-exponentials forms. Non-parametric estimation follows \cite{2014b,2011a}. Simulated event sequences are generated using Ogata's thinning method \cite{ogata1981thinning}.

\begin{figure}[!ht]
\centering
\begin{tikzpicture}[x=1cm, y=1cm, every node/.style={font=\small}, >=Latex]

\node (input) at (4, 12) [draw, rounded corners, minimum width=3.2cm, align=center] 
{Tick history \\ $(\tau - h, \tau]$};

\node (filter) at (2, 10.4) [draw, rounded corners, minimum width=3.2cm,align=center] 
{Filter: \\ $\mathcal{E} = \{T, C, PDM\}$};
\node (k1) at (2, 8.7) [draw, rounded corners, minimum width=3.2cm,align=center] 
{Estimate kernel \\ $\hat{\Phi}^{X,Y,r}$};
\node (sim1) at (2, 7.0) [draw, rounded corners, minimum width=3.2cm,align=center] 
{Simulate \\ $\tilde{N}^{X,Y,r}$};

\node (nofilter) at (6, 10.4) [draw, rounded corners, minimum width=3.2cm,align=center] 
{No filtering: \\ Use all ticks};
\node (k2) at (6, 8.7) [draw, rounded corners, minimum width=3.2cm,align=center] 
{Estimate kernel \\ $\hat{\Phi}^{X,Y}$};
\node (sim2) at (6, 7.0) [draw, rounded corners, minimum width=3.2cm,align=center] 
{Simulate \\ $\tilde{N}^{X,Y}$};

\node (rho) at (4, 5.0) [draw, thick, double, minimum width=4.4cm,align=center] 
{Compute arrival ratio \\ $\rho^{X,Y}_{(\tau, \tau + \xi]}$};

\node (decision) at (4, 3.2) [draw, thick, rounded corners, fill=gray!10, minimum width=4.4cm,align=center] 
{Reference leg decision \\ $\chi^{\mathcal{H}}(w)$};

\draw[->] (input) -- (filter);
\draw[->] (input) -- (nofilter);
\draw[->] (filter) -- (k1);
\draw[->] (k1) -- (sim1);
\draw[->] (nofilter) -- (k2);
\draw[->] (k2) -- (sim2);

\draw[->] (sim1) -- (rho);
\draw[->] (sim2) -- (rho);
\draw[->] (rho) -- (decision);

\end{tikzpicture}
\caption{Forecasting decision pipeline: kernel estimation, simulation, arrival ratio computation, and final decision $\chi^{\mathcal{H}}(w)$.}
\label{fig:hawkes_kernel_flow}
\end{figure}

\subsubsection{Forecast Evaluation and Model Selection}

To assess kernel performance, we compare simulated distributions \( \tilde{\mathcal{D}}^{X,m}_\omega \) of forecasted counts against realized counts \( N^{X,Y,r}_{(\tau,\tau+\xi]} \), computing the log-likelihood loss:
\begin{equation}
    \ell^{X,m} = -\sum_{j=1}^T \log \mathbb{P}\left( N^{X,Y,r}_{\omega_j} \mid \tilde{\mathcal{D}}^{X,m}_{\omega_j} \right)
\end{equation}

Model selection is based on the test for superior predictive ability (SPA) \cite{hansen2005test}, which compares competing kernels under common evaluation metrics while controlling for multiple testing.

The best-performing kernel \( \hat{\Phi}^{*} \) is used for all subsequent simulations. The resulting forecasting-based decision rule \( \chi^{\mathcal{H}}(w) \) is benchmarked against the oracle decision \( \chi^m(w) \) using the agreement score defined in Section~\ref{sec:problem_formulation}.

\begin{tcolorbox}[colback=white!95!gray, colframe=black!40!white, title=Summary: Integration into Evaluation Framework]
The Hawkes-based arrival ratio forecast yields a reference leg decision that anticipates short-term slippage. This decision is validated by comparing its alignment with the oracle decision across evaluation windows, allowing rigorous benchmarking against LOB- and tick-based quoting rules.
\end{tcolorbox}

\subsection{LOB-based quoting rule.}
Similarly, for each \( w \in \mathcal{W} \), we define a LOB-based quoting rule that uses the limit order book snapshot observed at time \( w_1 \) to evaluate expected slippage:
\begin{equation}
\mathcal{Q}^{\mathcal{B}}(w) := \arg\min_{r \in \{c,n\}} \mathbb{E}\left[\delta_S^{(r)}(w_2) \mid \mathcal{G}^{\mathcal{B}}(w)\right]
\end{equation}
with associated decision function:
\begin{equation}
\chi^{\mathcal{B}}(w) := \mathbb{I}\left\{ \mathcal{Q}^{\mathcal{B}}(w) = n \right\}
\end{equation}

These rules instantiate the framework in Section~\ref{sec:problem_formulation} using real-time market observables. The tick-based rule aggregates historical event data, and the LOB-based rule extracts instantaneous liquidity conditions. Both decisions use only information timestamped strictly before \(w_1\). The realized outcomes on \([w_1,w_2)\) are used solely for evaluation. When the two leg scores are exactly equal, we break ties in favor of \(c\) for determinism and comparability with the oracle.

\subsection{Reference Leg Selection via LOB Liquidity Profiles}
\label{sec:clf_forecasting}

Building on the previous section’s reference-leg selection rule derived from multivariate Hawkes intensity forecasts, we now introduce a LOB-state alternative that relies solely on the instantaneous limit order book configuration. Concretely, we aggregate level-wise depth with cross-level price gaps to form a composite liquidity profile (defined below). Whereas the Hawkes arrival ratio encodes temporal dependencies in event flow, this LOB-state construction encodes contemporaneous structure. We use this contrast to motivate a real-time, reference-leg selection rule based on liquidity discontinuities, with the empirical role of LOB-state information assessed against our evaluation oracle (deviation from the target spread).

Recent models of limit order book dynamics, including multivariate Hawkes formulations, offer detailed descriptions of event arrivals and state dependence \cite{chavez2012hawkes,protter2024queuehawkesmarkovian,mucciante2024orderbookhawkes}. In latency-sensitive HFT deployments, real-time estimation and updating can be computationally demanding, particularly as the number of event types and cross-excitation terms grows \cite{chavez2012hawkes}. This motivates a complementary rule that is computable at decision times from LOB-state features alone.

Motivated by the computational considerations in the preceding paragraph, we make explicit a practical trade-off between model expressiveness (rich event-history dependence) and runtime tractability (updates computable within each decision interval). In latency-sensitive deployments, low-latency decision rules with tractable state updates are commonly employed \cite{cartea2015algorithmic,avellaneda2008high}. Full re-estimation or online updating of high-dimensional event-history models can be computationally demanding \cite{chavez2012hawkes}. We therefore consider a complementary LOB-state framework that preserves LOB-aware structure while remaining tractable at runtime. As defined in Section~\ref{sec:methodology} and detailed in Section~\ref{sec:hawkes_forecasting}, the Hawkes arrival-ratio construction models the timing of events from history and treats the absolute price level as exogenous.

Grounding the tractability discussion above in a concrete LOB-state effect, cross-level price gaps can produce materially higher execution cost when available depth at the best quotes is insufficient \cite{cont2014price,bouchaud2002statistical}. In Table~\ref{tab:lob_t0}, the bid ladder exhibits a drop from \rupee 17450 at level 2 to \rupee 17401.1 at level 3, a \rupee 48.9 gap. An order that consumes the top two bid levels and then executes at level 3 would incur approximately \rupee 48.9 additional cost per contract due to this discontinuity. This motivates encoding price levels and cross-level gaps directly in our LOB-state measure for reference-leg selection, formalized in Section~\ref{sec:clf_forecasting}.

Extending the example above, the magnitude of cross-level price drops directly determines execution cost when top-of-book depth is insufficient, as illustrated in Table~\ref{tab:lob_t0}. Because our Hawkes arrival-ratio in Section~\ref{sec:hawkes_forecasting} treats the absolute price level as exogenous, it is not directly sensitive to these discontinuities. We therefore introduce a complementary LOB-state measure that encodes price levels, cross-level price gaps, and level-wise depth, and we assess its effect on reference-leg selection using the oracle benchmark defined in Section~\ref{sec:clf_forecasting}.

Building on the need to encode price levels, cross-level gaps, and depth stated above, the Composite Liquidity Factor (CLF) draws on established links between limit order book shape and execution cost. Hasbrouck’s \emph{log quote slope} (the rate of price change per unit depth) provides a core antecedent \cite{hasbrouck2001empirical}. Figure~\ref{fig:log_quote_slope_sym} depicts this slope as the gradient of the line connecting logarithmic quantity and price at the top levels; a greater quantity near the mid-price corresponds to a flatter slope and higher immediate liquidity. Prior work shows how shape and depth regularities affect short-horizon price impact and stability \cite{bouchaud2002statistical,cont2014price}, while Glosten’s decomposition of spreads links such shape effects to informational trading costs \cite{glosten1987components}. We adapt these ideas into a direction-sensitive LOB-state measure (CLF) for reference-leg selection.

\begin{figure}[h]
\centering
\begin{tikzpicture}[scale=0.8]

  \draw[->] (-3,0) -- (3.2,0) node[anchor=north, align=left] {Log\\quantity};
  \draw[->] (0,0) -- (0,4.2) node[anchor=east] {Price per share};

  \draw[thick] (0,0) -- (0,4);

  \draw[dashed, gray] (-1.75,1) -- (0,1);
  \draw[dashed, gray] (1.75,3.5) -- (0,3.5);
  \draw[dashed, gray] (-1.75,1) -- (-1.75,0);
  \draw[dashed, gray] (1.75,3.5) -- (1.75,0);

  \filldraw[black] (-1.75,1) circle (2pt) node[anchor=east] {\small $p^{b,i}$};
  \filldraw[black] (1.75,3.5) circle (2pt) node[anchor=west] {\small $p^{a,i}$};

  \draw[dashed, thick] (-1.75,1) -- (1.75,3.5);

  \node at (-1.75,-0.3) {\small $-\log(q^{b,i})$};
  \node at (1.75,-0.3) {\small $\log(q^{a,i})$};

\end{tikzpicture}
\caption{Illustration of the log quote slope: the slope of the dashed line connecting $( -\log(q^{b,i}), p^{b,i} )$ and $( \log(q^{a,i}), p^{a,i})$. Adapted from \cite{hasbrouck2001empirical}, Figure 2.}
\label{fig:log_quote_slope_sym}
\end{figure}

Although the \emph{log-quote slope} combines spread and depth across both sides of the LOB \cite{hasbrouck2001empirical}, reference-leg selection in our setting is side-conditional: we require a measure that emphasizes one-sided liquidity asymmetries across price gaps and level-wise depth on the side that will be consumed. The relevant side (bid or ask) is determined by whether the spread order is a buy or a sell. We therefore construct CLF features to encode these one-sided discontinuities (Section~\ref{sec:clf_forecasting}) and evaluate their contribution using deviation from the target spread.

To operationalise the side-conditional requirement above, we introduce a LOB-state measure that adapts the log-quote slope to one-sided execution. We refer to this measure as the Composite Liquidity Factor (CLF). CLF preserves the slope intuition while explicitly encoding price levels, cross-level price gaps, and level-wise depth on the side to be consumed; formal definitions follow in Section~\ref{sec:clf_forecasting}.

Let \(p^{Y,i}_X\) denote the price at level \(i\) on side \(Y\in\{b,a\}\) (bid or ask) for contract \(X\in\{c,n\}\) (current or next month), and let \(q^{Y,j}_X\) denote the available quantity at level \(j\). With the current-month contract used as the reference leg, we define a bid-side CLF at depth $i$ that contrasts the cross-level price gap with cumulative available depth on the consumed side:
\begin{equation}
  CLF^{b,i}_c
  \;=\;
  \frac{\log\!\big(p^{b,1}_c / p^{b,i+1}_c\big)}
       {\log\!\big(\sum_{j=1}^{i} q^{b,j}_c\big)},
  \qquad
  i \in \{1,\dots,\nu\!-\!1\},\;\; p^{b,i+1}_c < p^{b,i}_c .
\end{equation}
Here, the numerator encodes the price drop from the best bid to level $i\!+\!1$, and the denominator encodes cumulative bid-side depth through level $i$.

\noindent
Conversely, for the next-month contract used as the reference leg, we define an ask-side CLF at depth $i$ that contrasts the cross-level price gap with cumulative available depth on the consumed side:
\begin{equation}
  CLF^{a,i}_n
  \;=\;
  \frac{\log\!\big(p^{a,i+1}_n / p^{a,1}_n\big)}
       {\log\!\big(\sum_{j=1}^{i} q^{a,j}_n\big)},
  \qquad
  i \in \{1,\dots,\nu\!-\!1\},\;\; p^{a,i+1}_n > p^{a,i}_n .
\end{equation}
\noindent
Here, the numerator encodes the price increase from the best ask to level $i\!+\!1$, and the denominator encodes cumulative ask-side depth through level $i$.

This construction provides a flexible and direction-sensitive indicator of slippage risk, suited to the decision of reference contract selection in spread-based strategies. Unlike the Hawkes-based approach, the CLF-based rule requires no model fitting or simulation. All quantities used in the decision rule are computed directly from the current limit order book snapshot, making the method suitable for real-time application in latency-sensitive environments. 

We now define the CLF-based quoting rule formally, using the same decision function structure introduced in Section~\ref{sec:problem_formulation}.

\begin{tcolorbox}[colframe=black!60, colback=white!97!gray, boxrule=0.4pt, arc=2pt, left=4pt, right=4pt, top=4pt, bottom=4pt, width=\columnwidth]
\textbf{Definition: CLF-Based Reference Leg Selection Rule.}

\smallskip
\noindent
For each evaluation window \( w = [w_1, w_2) \in \mathcal{W} \), let \( \mathcal{G}^{\mathcal{B}}(w) \in \mathbb{R}^2 \) denote the LOB-based filter output containing directional liquidity values for the two contract legs.
 Define:
\begin{align*}
&\mathcal{G}^{\mathcal{B}}(w) := \left( CLF_c(w),\; CLF_n(w) \right) \\
&\mathcal{Q}^{\mathcal{B}}(w) := \arg\min_{r \in \{c,n\}} CLF_r(w) \\
&\chi^{\mathcal{B}}(w) := \mathbb{I} \left\{ \mathcal{Q}^{\mathcal{B}}(w) = n \right\}
\end{align*}
where \( CLF_r(w) \) denotes the composite liquidity factor for leg \( r \in \{c, n\} \), and \( n \) refers to the next-month contract.
\noindent
\end{tcolorbox}

This decision rule operates entirely on the LOB-based filtered representation \( \mathcal{G}^{\mathcal{B}}(w) \), requiring no historical data, simulation, or model fitting. It is fully deterministic and computed from observable execution risk at the quoting time \( w_1 \). By capturing asymmetries in directional liquidity across contract legs, \( \mathcal{G}^{\mathcal{B}}(w) \) provides a structurally simpler, real-time quoting signal that complements event-driven methods.

The two approaches discussed in this paper reflect different assumptions about the information content of market data. The Hawkes-based rule treats reference-impacting events as informative flow variables, modeling their interactions over time. In contrast, the CLF-based rule assumes that execution-relevant information is embedded in the instantaneous structure of the limit order book.

\section{Data}
This section describes the tick-driven market data and the microstructural features used by our filters. We use tick-by-tick (TBT) data from the National Stock Exchange (NSE) of India; the experiment analyzes NIFTY futures on February 7, 2022.\footnote{All exchange specifications and market microstructure rules referenced in this study reflect those in effect as of March 2022, corresponding to the period from which the data were collected.}
The dataset includes two instruments: the current-month contract expiring on February 24, 2022 (\(F_c\)), and the next-month contract expiring on March 31, 2022 (\(F_n\)). As shown in Table~\ref{tab:tick_counts}, both contracts display substantial tick activity on the sample date, providing granularity for execution modeling.

Each tick is one of four event types: \texttt{NEW} (order addition), \texttt{MODIFY} (order revision), \texttt{CANCEL} (order withdrawal), and \texttt{TRADE} (execution). These types are used to extract features relevant to quoting behavior and limit order book dynamics. The minimum tick size for NIFTY futures is \rupee 0.05, as mandated by NSE.

In our methodology, this data is segmented into a collection of evaluation windows \( \mathcal{W} = \{w = [w_1, w_2) \} \), where each window spans a short time interval (e.g., 100 milliseconds) aligned with the quoting decision frequency. For each \( w \in \mathcal{W} \), we apply two distinct classes of filters:

\begin{enumerate}[label={}]
  \item The tick-based filter \( \mathcal{G}^{\mathcal{T}}(w) \) operates on the order flow observed strictly before the quoting time \( w_1 \).
  \item The LOB-based filter \( \mathcal{G}^{\mathcal{B}}(w) \) extracts features from the limit order book snapshot observed exactly at time \( w_1 \).
\end{enumerate}

These filters produce real-valued representations of prevailing market conditions that serve as inputs to the quoting rules defined in Section~\ref{sec:problem_formulation}.

\paragraph{Tick-derived features.}
For each evaluation window \( w \in \mathcal{W} \), the tick-based filter \( \mathcal{G}^{\mathcal{T}}(w) \) encodes summary statistics of the recent order flow observed strictly before the quoting time \( w_1 \). The features chosen as filtration criteria are those event types that worsen the price stability of the selected reference contract. These features include:
\begin{enumerate}[label={}]
    \item Count of executed trades (\texttt{TRADE} events).
    \item Count of order cancellations (\texttt{CANCEL} events).
    \item Count of price-worsening quote updates (\texttt{MODIFY} events where the best bid decreases or the best ask increases).
    \item Total number of tick events, inclusive of all above types.
\end{enumerate}

\paragraph{LOB-derived features.}
Analogous to tick-derived features, we define features derived from the current limit order book (LOB) to evaluate reference-leg selection. In each interval \( w \in \mathcal{W} \), the LOB-based filter \( \mathcal{G}^{\mathcal{B}}(w) \) extracts directional liquidity indicators from the limit order book snapshot observed at time \( w_1 \). These features are computed separately for each contract leg and for each relevant side of the LOB (bid or ask), based on whether the leg is expected to be lifted or hit under the quoting convention.

Specifically, we extract:
\begin{enumerate}[label={}]
    \item Price at each level \( p^{b,i}, p^{a,i} \), for \( i = 1, \dots, \nu \)
    \item Quantity available at each level \( q^{b,i}, q^{a,i} \), for \( i = 1, \dots, \nu \)
\end{enumerate}

These level-wise features are later used to construct directional liquidity measures such as the Composite Liquidity Factor (CLF), introduced in a subsequent section, which serve as inputs to the LOB-based quoting rule \( \mathcal{Q}^{\mathcal{B}} \).

\section{Experiment and Results}

From a market microstructure perspective, the two decision rules are designed to respond to different aspects of market information. The CLF is computed directly from limit order book primitives: instantaneous depth, queue imbalance, and local LOB shape, which are widely used as proxies for short-horizon liquidity and price impact in limit-order markets \cite{avellaneda2008high,cont2014price}. This construction makes the CLF-based rule sensitive to rapid changes in the relative liquidity of the two legs over short horizons. However, during episodes of intense trading activity, high-frequency quotes and trades are heavily affected by microstructure noise and fleeting orders. Hasbrouck and Saar \cite{hasbrouck2013low} document that low-latency trading can amplify such transitory LOB dynamics. In such environments, an exclusively cross-sectional, LOB-based statistic can generate spurious or very short-lived signals.

The Hawkes arrival-ratio rule, in contrast, is a low-dimensional summary of the event-time dynamics of price-impacting messages. It aggregates information about how often price-impacting events arrive on each leg, with an explicit memory structure. Consistent with empirical evidence that Hawkes-based event-time models capture clustering and persistent order-flow dynamics~\cite{bacry2015hawkes,hardiman2013predicting,chavez2012hawkes}, we expect this construction to be relatively insensitive to purely transitory fluctuations in the order-flow and better suited to detecting and exploiting persistent regimes in the benchmark.

We deliberately differentiate the information sets of the two rules. For the Hawkes arrival-ratio we use only \emph{reference-impacting} events at the top of the book (arrivals, cancellations, and downward price modifications that change the best quotes). For the CLF-based rule, we construct LOB-state measures from the first five depth levels on each side, thereby capturing the local cross-sectional shape of the LOB. In this sense, the comparison pits a history of reference-impacting ticks against a smoothed history of LOB snapshots. Our empirical analysis focuses on February 7, 2022 and on NIFTY futures expiring in February (\(F_c\)) and March (\(F_n\)). Throughout, the Hawkes arrival-ratio rule and the CLF-based LOB-state rule are evaluated within the decision-function framework of Section~\ref{sec:problem_formulation}, using the oracle \( \chi^m(w) \) as a benchmark. Table~\ref{tab:tick_counts} reports tick-event counts for each contract on February 7, 2022. Counts include all four TBT event types (\texttt{NEW}, \texttt{MODIFY}, \texttt{CANCEL}, \texttt{TRADE}). The substantial volume supports evaluation of quoting policies at high temporal resolution.

\begin{table}[!ht]
    \centering
    \small
    \begin{tabular}{lll}
    \toprule
         \textbf{Contract} & \textbf{Date} & \textbf{Tick events (count)} \\
    \midrule
        NIFTY February Futures & 2022-02-07 & 49,828,303 \\
        NIFTY March Futures    & 2022-02-07 & 16,349,197 \\
    \bottomrule
    \end{tabular}
    \caption{Tick-event counts by contract on February 7, 2022. Counts include \texttt{NEW}, \texttt{MODIFY}, \texttt{CANCEL}, and \texttt{TRADE} events. A sample from the processed tick data is provided in the appendix, in Table~\ref{tab:trade_data} for illustration.}
    \label{tab:tick_counts}
\end{table}

The evaluation period contains both short-lived bursts in which the oracle reference leg flips frequently and extended segments in which the oracle remains on the same leg for many consecutive windows. The presence of both low- and high-persistence benchmark regimes within a single day is useful for our purposes: it allows us to examine how order-flow based and LOB-based rules behave under different benchmark regimes while keeping the market context fixed. Among the runs, Figure~\ref{fig:benchmark_persistence_runs} shows that benchmark regimes are distributed almost symmetrically across the two legs: across the four bins \(Q_1\)–\(Q_4\) the oracle generates 124 runs on \(F_c\) and 123 runs on \(F_n\). This symmetry emphasises that benchmark regimes are not dominated by a single contract in terms of the number of episodes. In the subsequent analysis we investigate this structure by conditioning agreement scores on the run-length and oracle's benchmark contract, so that the relative performance of the Hawkes arrival-ratio and CLF can be interpreted in terms of both the duration and the contract choice of benchmark regimes.

\begin{figure}[!ht]
  \centering
  \begin{tikzpicture}[x=2.0cm,y=0.039cm,>=latex,
                      every node/.style={font=\footnotesize}]
    \definecolor{nearCol}{RGB}{102,194,165}   
    \definecolor{farCol}{RGB}{141,160,203}    

    \def\RQoneNear{33}
    \def\RQoneFar{38}
    \def\RQtwoNear{29}
    \def\RQtwoFar{23}
    \def\RQthreeNear{32}
    \def\RQthreeFar{30}
    \def\RQfourNear{30}
    \def\RQfourFar{32}

    \def\RQoneTot{71}
    \def\RQtwoTot{52}
    \def\RQthreeTot{62}
    \def\RQfourTot{62}

    \draw[->] (0,0) -- (0,85) node[above] {Number of runs};
    \draw[->] (0,0) -- (5.0,0);

    \foreach \y in {0,20,40,60,80} {
      \draw[densely dotted,gray!60] (0,\y) -- (5.0,\y);
      \node[left] at (0,\y) {\footnotesize \y};
    }

    \def\xQone{1.0}
    \def\xQtwo{2.0}
    \def\xQthree{3.0}
    \def\xQfour{4.0}
    \def\barW{0.35}

    \draw[fill=nearCol,draw=black]
      ({\xQone-\barW},0) rectangle ({\xQone+\barW},\RQoneNear);
    \draw[fill=farCol,draw=black]
      ({\xQone-\barW},\RQoneNear) rectangle ({\xQone+\barW},{\RQoneNear+\RQoneFar});
    \node[above] at (\xQone,\RQoneTot) {\scriptsize 71};
    \node at (\xQone,16.5) {\scriptsize 33};
    \node at (\xQone,52)   {\scriptsize 38};

    \draw[fill=nearCol,draw=black]
      ({\xQtwo-\barW},0) rectangle ({\xQtwo+\barW},\RQtwoNear);
    \draw[fill=farCol,draw=black]
      ({\xQtwo-\barW},\RQtwoNear) rectangle ({\xQtwo+\barW},{\RQtwoNear+\RQtwoFar});
    \node[above] at (\xQtwo,\RQtwoTot) {\scriptsize 52};
    \node at (\xQtwo,14.5) {\scriptsize 29};
    \node at (\xQtwo,40.5) {\scriptsize 23};

    \draw[fill=nearCol,draw=black]
      ({\xQthree-\barW},0) rectangle ({\xQthree+\barW},\RQthreeNear);
    \draw[fill=farCol,draw=black]
      ({\xQthree-\barW},\RQthreeNear) rectangle ({\xQthree+\barW},{\RQthreeNear+\RQthreeFar});
    \node[above] at (\xQthree,\RQthreeTot) {\scriptsize 62};
    \node at (\xQthree,16) {\scriptsize 32};
    \node at (\xQthree,47) {\scriptsize 30};

    \draw[fill=nearCol,draw=black]
      ({\xQfour-\barW},0) rectangle ({\xQfour+\barW},\RQfourNear);
    \draw[fill=farCol,draw=black]
      ({\xQfour-\barW},\RQfourNear) rectangle ({\xQfour+\barW},{\RQfourNear+\RQfourFar});
    \node[above] at (\xQfour,\RQfourTot) {\scriptsize 62};
    \node at (\xQfour,15) {\scriptsize 30};
    \node at (\xQfour,46) {\scriptsize 32};

    \node[below=8pt,align=center] at (\xQone,0)
      {$Q_1$\\(1--3 windows)};
    \node[below=8pt,align=center] at (\xQtwo,0)
      {$Q_2$\\(4--10 windows)};
    \node[below=8pt,align=center] at (\xQthree,0)
      {$Q_3$\\(11--58 windows)};
    \node[below=8pt,align=center] at (\xQfour,0)
      {$Q_4$\\($\geq 59$ windows)};

    \draw[fill=nearCol,draw=black] (4.5,74) rectangle (4.8,77);
    \node[anchor=west] at (4.85,75.5) {Oracle selects $F_c$};

    \draw[fill=farCol,draw=black] (4.5,68) rectangle (4.8,71);
    \node[anchor=west] at (4.85,69.5) {Oracle selects $F_n$};

  \end{tikzpicture}

  \caption{Number of oracle runs in each benchmark-persistence bin on February 7, 2022, split by the benchmark’s selected reference leg.}
  \label{fig:benchmark_persistence_runs}
\end{figure}

\begin{figure}[!ht]
  \centering
  \begin{tikzpicture}[x=2.0cm,y=0.039cm,>=latex,
                      every node/.style={font=\footnotesize}]
    \definecolor{qoneCol}{RGB}{222,235,247}   
    \definecolor{qtwoCol}{RGB}{189,215,231}
    \definecolor{qthreeCol}{RGB}{107,174,214}
    \definecolor{qfourCol}{RGB}{33,113,181}   

    \def\xQone{1.0}
    \def\xQtwo{2.0}
    \def\xQthree{3.0}
    \def\xQfour{4.0}
    \def\barW{0.35}

    \draw[->] (0,0) -- (0,105) node[above] {Share of windows (\%)};
    \draw[->] (0,0) -- (5.0,0);

    \foreach \y in {0,20,40,60,80,100} {
      \draw[densely dotted,gray!60] (0,\y) -- (5.0,\y);
      \node[left] at (0,\y) {\footnotesize \y};
    }

    \def\Sone{0.38}
    \def\Stwo{0.91}
    \def\Sthree{4.64}
    \def\Sfour{94.07}

    \draw[fill=qoneCol!90,draw=black]
      ({\xQone-\barW},0) rectangle ({\xQone+\barW},\Sone);
    \draw[fill=qtwoCol!90,draw=black]
      ({\xQtwo-\barW},0) rectangle ({\xQtwo+\barW},\Stwo);
    \draw[fill=qthreeCol!90,draw=black]
      ({\xQthree-\barW},0) rectangle ({\xQthree+\barW},\Sthree);
    \draw[fill=qfourCol!90,draw=black]
      ({\xQfour-\barW},0) rectangle ({\xQfour+\barW},\Sfour);

    \node[above,align=center] at (\xQone,\Sone)
      {\scriptsize 0.4\%};
    \node[above,align=center] at (\xQtwo,\Stwo)
      {\scriptsize 0.9\%};
    \node[above,align=center] at (\xQthree,\Sthree)
      {\scriptsize 4.6\%};
    \node[above,align=center] at (\xQfour,\Sfour)
      {\scriptsize 94.1\%};

    \node[below=6pt,align=center] at (\xQone,0) {$Q_1$};
    \node[below=6pt,align=center] at (\xQtwo,0) {$Q_2$};
    \node[below=6pt,align=center] at (\xQthree,0) {$Q_3$};
    \node[below=6pt,align=center] at (\xQfour,0) {$Q_4$};
  \end{tikzpicture}

  \caption{Share of evaluation windows in each benchmark-persistence bin on February 7, 2022.}
  \label{fig:benchmark_persistence_windows}
\end{figure}

To study how performance varies with the stability of the oracle, we construct what we call \emph{benchmark persistence}. Starting from the sequence of oracle decisions \( \{\chi^m(w)\}_w \), we segment the evaluation period into maximal contiguous runs during which the oracle selects the same reference leg. The length of a run, measured in evaluation windows, is our measure of persistence: very short runs correspond to rapid benchmark switches between contracts, whereas long runs correspond to extended periods in which the oracle consistently prefers the same leg.

We then examine the empirical distribution of run lengths on February 7, 2022 and partition runs into \emph{bins of benchmark persistence}. The first bin contains runs of length 1--3 windows, the second bin contains runs of length 4--10 windows, the third bin collects runs of length 11--58 windows, and the fourth bin consists of runs of length at least 59 windows. Figures~\ref{fig:benchmark_persistence_runs} and~\ref{fig:benchmark_persistence_windows} summarise this structure. Figure~\ref{fig:benchmark_persistence_runs} reports the number of oracle runs in each persistence bin, while Figure~\ref{fig:benchmark_persistence_windows} shows the share of evaluation windows that fall into each bin. The first two bins account for many distinct runs (72 and 52, respectively) but together cover only about \( 1.3\% \) of evaluation windows. The third bin contains 62 medium-length runs and accounts for roughly \( 4.6\% \) of windows. The fourth bin also contains 62 runs, but these long, high-persistence regimes account for about \( 94\% \) of the evaluation period.

Splitting runs by the oracle’s preferred contract shows that benchmark persistence is driven more by regime duration than by regime count. Across all four bins \(Q_1\)–\(Q_4\), the numbers of runs on the two contracts are similar. Over the full evaluation period, however, the oracle selects the near leg in about \(72\%\) of windows. Within the highest-persistence bin \(Q_4\), runs on the near leg last on average roughly three times as long as runs on the other leg and together account for close to \(69\%\) of all windows. By contrast, the low- and medium-persistence bins \(Q_1\)–\(Q_3\) cover only about \(6\%\) of the day and are more evenly split across the two contracts. Benchmark regimes are therefore concentrated in long intervals during which one leg remains market-optimal, punctuated by shorter excursions in which the oracle switches to the other leg.

\subsection{Hawkes-Based Forecasting.}
Within each decision interval \(w\), the rule \(\chi^{\mathcal{H}}(w)\) in Section~\ref{sec:hawkes_forecasting} uses simulated forward intensities from a multivariate Hawkes model. Using only data available at the start of \(w\), we estimate the Hawkes kernels over the past 1000\,ms for the reference-impacting events by maximum likelihood, then simulate \(K\) forward paths over the next 10\,ms. From the simulated reference-side event counts, we compute arrival ratios \(\rho^{X,Y}_{(w)}\) and apply the prescribed threshold to issue the binary reference-leg decision \(\chi^{\mathcal{H}}(w)\in\{0,1\}\).

\subsection{CLF-Based Rule.}
At each tick, we compute CLF at depths 1--4 for both \(F_c\) and \(F_n\) \((CLF_1,\dots,CLF_4)\). CLF contrasts the cross-level price gap with cumulative depth, so that a lower value indicates more favourable immediate liquidity on the consumed side. The contract with the lower per-tick CLF score is selected at that tick as the reference contract. We partition the evaluation period into decision windows \(w\) of length \(10\,\text{ms}\), shifted forward in \(10\,\text{ms}\) steps. For each window \(w\), we take a majority vote over the tick-level reference choices observed in the preceding \(1{,}000\,\text{ms}\). This yields the reference-leg decision \(\chi^{\mathcal{B}}(w)\) applied to the next \(10\,\text{ms}\) interval, with ties broken in favour of the next-month contract \(F_n\).

\begin{tcolorbox}[colback=gray!3, colframe=black!30, title=Modeling Remark]
Tick-level CLF values must be mapped to the interval-based decision framework of Section~\ref{sec:problem_formulation}. We construct the LOB-based filter \( \mathcal{G}^{\mathcal{B}} \) by applying an exponential moving average (EMA) to per-tick CLF observations over the historical window \( (w-\Delta,\,w) \), using only information available before the start of the interval. Let \(c_t\) denote the CLF at tick time \(t<w\). In our implementation we set \(\Delta = 1{,}000\,\text{ms}\), so that the EMA smooths CLF over the same 1{,}000\,ms history used in the majority-vote construction and in the Hawkes-based rule’s estimation window.
\end{tcolorbox}

\subsection{Market-Optimal Benchmark.}
Before computing the Hawkes-based rule and the CLF-based rule, we use the method presented in Section~\ref{sec:market_benchmark} for evaluation. We construct a reference oracle \( \chi^m(w) \in \{0,1\} \), which selects the contract that would have yielded the lower realized spread in hindsight. For each window \( w \in \mathcal{W} \), we compute the actual execution spreads for both reference leg choices and select the contract that minimizes slippage.

\subsection{Evaluation Metric.}
Each method’s prediction accuracy is measured using the agreement score \( \mathcal{A}(\chi, \chi^m) \), as defined in Section~\ref{sec:problem_formulation}:
\[
\mathcal{A}(\chi, \chi^m) := \frac{1}{|\mathcal{W}|} \sum_{w \in \mathcal{W}} \mathbb{I} \left\{ \chi(w) = \chi^m(w) \right\}
\]
This metric reflects how often each rule’s decision matches the market-optimal choice based on realized execution outcomes. To contextualize the empirical comparison, we summarize the three decision rules evaluated in this experiment.

\begin{tcolorbox}[colframe=black!60, colback=white!97!gray, boxrule=0.4pt, arc=2pt,
left=4pt, right=4pt, top=4pt, bottom=4pt, width=\columnwidth]
\textbf{Summary of Reference Leg Decision Rules.}
\smallskip
\textbf{Hawkes-based rule:} \\ \( \chi^{\mathcal{H}}(w) = \mathbb{I}\left\{ \rho^{c,a}(w) \geq \rho^{n,b}(w) \right\} \) \\
\textbf{CLF-based rule:} \\ \( \chi^{\mathcal{B}}(w) = \mathbb{I}\left\{ CLF_n(w) < CLF_c(w) \right\} \) \\
\textbf{Market-optimal rule:} \\ \( \chi^m(w) = \arg\min_{r \in \{c, n\}} \delta_S^{(r)}(w) \)
\end{tcolorbox}

This evaluation framework enables a direct comparison of flow-based and state-based decision models under realistic market conditions, using slippage-minimizing outcomes as ground truth.

We divide the results into two parts. First, we compare alternative Hawkes kernel specifications using the test for superior predictive ability \cite{hansen2005test}. Second, we evaluate the reference-leg selection rules, \(\chi^{\mathcal{H}}(w)\) and \(\chi^{\mathcal{B}}(w)\), using deviation from the target spread, with both benchmarked against the oracle \(\chi^{m}(w)\) introduced in Section~\ref{sec:problem_formulation}.

\subsection{Hawkes Kernel Selection}

Table~\ref{table:hawkes_results} reports SPA \cite{hansen2005test} test values for four kernel specifications: two parametric (HawkesExp and HawkesSumExp) and two non-parametric (HawkesEM and HawkesCondLaw). The parametric methods yield comparable test values of 0.017 and 0.015, respectively, indicating similar performance. The non-parametric methods exhibit greater variability: the EM-based kernel records a substantially higher test value of 0.530, while the conditional law kernel yields 0.0. The results suggest that the expectation-maximization kernel \cite{2011a} is not outperformed by any of the other candidate models when used as a benchmark.

\begin{table}[!ht]
    \centering
    \caption{SPA test values for Hawkes kernel specifications}
	\begin{tabular}{@{}llll@{}}
        \toprule
		HawkesExp & HawkesSumExp & HawkesEM & HawkesCondLaw \\
        \midrule
		0.017     & 0.015         & 0.530    & 0.000 \\
        \bottomrule
	\end{tabular}
    \label{table:hawkes_results}
\end{table}

\paragraph{SPA setup and interpretation.}
For each candidate specification \(k\), we treat \(k\) as the benchmark and test whether any competing model achieves a lower expected loss than \(k\) under the chosen loss function. We report the SPA \emph{p}-value \(p_k\) for this benchmarked comparison. Small values of \(p_k\) lead to rejection of the null (there is evidence that at least one competing model outperforms the benchmark \(k\)); large values of \(p_k\) imply we fail to reject the null (no evidence, in this sample, that any competitor is superior to benchmark \(k\)). Accordingly, higher \emph{p}-values do not establish optimality, but they indicate that the chosen benchmark withstands the SPA comparison on this sample~\cite{hansen2005test}.

Since the SPA test does not reject the null hypothesis that the EM-based kernel is the benchmark, we adopt the HawkesEM specification for all subsequent analyses. Consistent with evidence that high-frequency order-flow intensities exhibit time-of-day patterns~\cite{omi2017timedep}, we report intraday kernel variation, as shown in Figures~\ref{fig:excitation-grid}.

\subsection{Comparison of Hawkes Arrival Ratio and CLF}
\label{subsec:hawkes_clf_comparison}

Using the forecasting rules defined in the preceding subsections and the agreement metric \( \mathcal{A}(\chi, \chi^m) \) introduced in Section~\ref{sec:problem_formulation}, we compare the Hawkes-based rule \( \chi^{\mathcal{H}}(w) \) and the CLF-based rule \( \chi^{\mathcal{B}}(w) \) against the market-optimal oracle \( \chi^m(w) \). The evaluation uses high-frequency tick data from February 7, 2022 for the NIFTY futures contracts \( F_c \) and \( F_n \). All CLF parameters and Hawkes kernel coefficients are estimated on a training sample; the evaluation period is used only for generating forecasts and for computing agreement scores. The oracle benchmark \( \chi^m(w) \) is the only object that uses realised post-decision information when defining the market-optimal reference leg.

\begin{table}[!ht]
  \centering
  \caption{Agreement score \( \mathcal{A}(\chi, \chi^m) \): proportion of evaluation windows in which each method’s predicted reference leg matches the market-optimal oracle on February 7, 2022. The joint-agreement row reports the fraction of windows in which both the Hawkes arrival ratio and CLF\textsubscript{4} simultaneously match the oracle.}
  \label{tab:clf_hawkes_results}
  \begin{tabular}{lc}
    \toprule
    \textbf{Method} & \textbf{Agreement score} \\
    \midrule
    CLF\textsubscript{1} & 0.3796 \\
    CLF\textsubscript{2} & 0.4754 \\
    CLF\textsubscript{3} & 0.5919 \\
    CLF\textsubscript{4} & 0.6358 \\
    Hawkes arrival ratio & \textbf{0.6856} \\
    Hawkes--CLF\textsubscript{4} joint agreement & \textbf{0.4149} \\
    \bottomrule
  \end{tabular}
\end{table}

Table~\ref{tab:clf_hawkes_results} shows that, on this evaluation day, the Hawkes arrival-ratio rule attains an agreement score of \( 0.6856 \) with the market-optimal oracle. Among the CLF variants, CLF\textsubscript{4} achieves the highest agreement score at \( 0.6358 \), followed by CLF\textsubscript{3}, CLF\textsubscript{2}, and CLF\textsubscript{1}. On February 7, 2022, therefore, the low-dimensional Hawkes arrival ratio extracted from the limit-order dynamics attains a modestly higher agreement with the oracle than the best-performing CLF specification.

To quantify how often the two rules arrive at the same decision, Figure~\ref{fig:hawkes_clf4_joint} and the joint-agreement row of Table~\ref{tab:clf_hawkes_results} report the fraction of evaluation windows in which each combination of outcomes occurs. In \( 41.49\% \) of windows, both the Hawkes arrival ratio and CLF\textsubscript{4} match the oracle. The Hawkes arrival ratio is uniquely in agreement with the oracle in \( 27.08\% \) of windows, while CLF\textsubscript{4} is uniquely in agreement in \( 22.10\% \). Only \( 9.34\% \) of windows are misclassified by both rules simultaneously.

Conditional on the Hawkes arrival ratio agreeing with the oracle, CLF\textsubscript{4} is also in agreement in about \( 60.5\% \) of windows; conditional on CLF\textsubscript{4} agreeing with the oracle, the Hawkes arrival ratio is also in agreement in about \( 65.2\% \). These conditional shares indicate substantial overlap between the two rules, together with a non-trivial set of windows in which only one of them matches the oracle. A hypothetical aggregator that could choose, ex ante, the better of the two signals in each window would miss the benchmark leg in at most the \( 9.34\% \) of windows where both rules disagree with the oracle.

Panels~(b) and~(c) of Figure~\ref{fig:hawkes_clf4_joint} separate agreement by the oracle’s selected contract for each rule in isolation. In these panels, the rows distinguish evaluation windows in which the oracle selects the current-month contract \(F_c\) from those in which it selects the next-month contract \(F_n\), while the columns indicate whether the rule matches or deviates from the oracle. Panel~(b) shows that the Hawkes arrival-ratio rule agrees with the oracle in approximately \(88\%\) of windows when the oracle selects \(F_c\), but only in about \(18\%\) of windows when the oracle selects \(F_n\). Panel~(c) shows that CLF\textsubscript{4} attains agreement of about \(61\%\) when the oracle selects \(F_c\) and about \(69\%\) when it selects \(F_n\). Taken together, these contract-wise panels indicate that the Hawkes arrival ratio is most informative in regimes where the current-month contract is the market-optimal reference leg, whereas CLF\textsubscript{4} contributes relatively more information when the oracle favours the next-month contract.

\begin{figure}[!htbp]
  \centering
  \begin{tikzpicture}[x=2.0cm,y=2.0cm,>=latex,
                      every node/.style={font=\footnotesize}]
    \definecolor{bothagree}{RGB}{102,194,165} 
    \definecolor{hawkesonly}{RGB}{57,106,177} 
    \definecolor{clfonly}{RGB}{218,124,48}    
    \definecolor{bothdis}{RGB}{204,76,76}     

    \begin{scope}[shift={(0,2.6)}]
      \draw[thick] (0,0) rectangle (2,2);
      \draw[thick] (1,0) -- (1,2);
      \draw[thick] (0,1) -- (2,1);

      \path[fill=bothagree!60] (0,0) rectangle (1,1); 
      \path[fill=hawkesonly!40] (1,0) rectangle (2,1); 
      \path[fill=clfonly!40]  (0,1) rectangle (1,2);   
      \path[fill=bothdis!40]   (1,1) rectangle (2,2);  

      \node at (0.5,0.5) {\shortstack{41.49\%\\[0.1em](14{,}502)}};
      \node at (1.5,0.5) {\shortstack{27.08\%\\[0.1em](9{,}465)}};
      \node at (0.5,1.5) {\shortstack{22.10\%\\[0.1em](7{,}724)}};
      \node at (1.5,1.5) {\shortstack{9.34\%\\[0.1em](3{,}265)}};

      \node[below=4pt] at (0.5,0) {agree};
      \node[below=4pt] at (1.5,0) {disagree};
      \node[below=10pt,font=\footnotesize\bfseries] at (1,0) {CLF\textsubscript{4}};

      \node[left=4pt] at (0,0.5) {agree};
      \node[left=4pt] at (0,1.5) {disagree};
      \node[rotate=90,left=12pt,font=\footnotesize\bfseries] at (-0.3,1.275) {Hawkes};

      \node[font=\footnotesize\bfseries] at (1,2.25) {(a) All windows};
    \end{scope}

    \begin{scope}[shift={(2.8,2.6)}]
      \draw[thick] (0,0) rectangle (2,2);
      \draw[thick] (1,0) -- (1,2);
      \draw[thick] (0,1) -- (2,1);


      \path[fill=bothagree!60] (0,1) rectangle (1,2);
      \path[fill=bothdis!40]   (1,1) rectangle (2,2);
      \path[fill=bothagree!60] (0,0) rectangle (1,1);
      \path[fill=bothdis!40]   (1,0) rectangle (2,1);

      \node at (0.5,1.5) {\shortstack{88.18\%\\[0.1em](22{,}238)}};
      \node at (1.5,1.5) {\shortstack{11.82\%\\[0.1em](2{,}981)}};
      \node at (0.5,0.5) {\shortstack{17.76\%\\[0.1em](1{,}729)}};
      \node at (1.5,0.5) {\shortstack{82.24\%\\[0.1em](8{,}007)}};

      \node[below=4pt] at (0.5,0) {agree};
      \node[below=4pt] at (1.5,0) {disagree};
      \node[below=10pt,font=\footnotesize\bfseries] at (1,0) {Hawkes};

      \node[left=4pt] at (0,0.5) {$F_n$};
      \node[left=4pt] at (0,1.5) {$F_c$};
      \node[rotate=90,left=12pt,font=\footnotesize\bfseries] at (-0.3,1.275) {Oracle leg};

      \node[font=\footnotesize\bfseries] at (1,2.25) {(b) Hawkes by contract};
    \end{scope}

    \begin{scope}[shift={(0,-0.2)}]
      \draw[thick] (0,0) rectangle (2,2);
      \draw[thick] (1,0) -- (1,2);
      \draw[thick] (0,1) -- (2,1);


      \path[fill=bothagree!60] (0,1) rectangle (1,2);
      \path[fill=bothdis!40]   (1,1) rectangle (2,2);
      \path[fill=bothagree!60] (0,0) rectangle (1,1);
      \path[fill=bothdis!40]   (1,0) rectangle (2,1);

      \node at (0.5,1.5) {\shortstack{61.33\%\\[0.1em](15{,}468)}};
      \node at (1.5,1.5) {\shortstack{38.67\%\\[0.1em](9{,}751)}};
      \node at (0.5,0.5) {\shortstack{69.41\%\\[0.1em](6{,}758)}};
      \node at (1.5,0.5) {\shortstack{30.59\%\\[0.1em](2{,}978)}};

      \node[below=4pt] at (0.5,0) {agree};
      \node[below=4pt] at (1.5,0) {disagree};
      \node[below=10pt,font=\footnotesize\bfseries] at (1,0) {CLF\textsubscript{4}};

      \node[left=4pt] at (0,0.5) {$F_n$};
      \node[left=4pt] at (0,1.5) {$F_c$};
      \node[rotate=90,left=12pt,font=\footnotesize\bfseries] at (-0.3,1.275) {Oracle leg};

      \node[font=\footnotesize\bfseries] at (1,2.25) {(c) CLF\textsubscript{4} by contract};
    \end{scope}

    \begin{scope}[shift={(2.8,1.1675)}]
      \node[anchor=west,font=\footnotesize\bfseries] at (-0.2,0.9) {Outcome};
      \draw[fill=bothagree!60,draw=black] (0,0.35) rectangle (0.3,0.65);
      \node[anchor=west] at (0.35,0.50) {Agreement};
      \draw[fill=hawkesonly!40,draw=black] (0,0.05) rectangle (0.3,0.35);
      \node[anchor=west] at (0.35,0.20) {Hawkes only (panel a)};
      \draw[fill=clfonly!40,draw=black] (0,-0.25) rectangle (0.3,0.05);
      \node[anchor=west] at (0.35,-0.10) {CLF\textsubscript{4} only (panel a)};
      \draw[fill=bothdis!40,draw=black] (0,-0.55) rectangle (0.3,-0.25);
      \node[anchor=west] at (0.35,-0.40) {Disagreement};
    \end{scope}

  \end{tikzpicture}
  \caption{Joint agreement of the Hawkes arrival-ratio rule and CLF\textsubscript{4} relative to the oracle on February 7, 2022. Panel~(a) reports unconditional shares of evaluation windows in each joint outcome. Panel~(b) reports contract-wise agreement and disagreement of the Hawkes arrival-ratio rule with the oracle. Panel~(c) reports contract-wise agreement and disagreement of CLF\textsubscript{4} with the oracle. Each cell reports the percentage and the number of evaluation windows corresponding to that outcome, using a common colour palette across panels.}
  \label{fig:hawkes_clf4_joint}
\end{figure}

Given the partition in~\ref{fig:benchmark_persistence_runs}, each evaluation window inherits the benchmark-persistence bin of the run to which it belongs. Figure~\ref{fig:benchmark_persistence_accuracy} reports agreement scores for the Hawkes arrival-ratio and CLF\textsubscript{4} within each persistence bin. In the lowest-persistence bin (runs of length 1--3 windows), CLF\textsubscript{4} attains an agreement score of approximately \( 0.459 \), compared with \( 0.406 \) for the Hawkes arrival ratio. In the second bin (runs of length 4--10 windows), CLF\textsubscript{4} reaches an agreement score of \( 0.524 \) and the Hawkes arrival ratio reaches \( 0.489 \). In the third bin (runs of length 11--58 windows), CLF\textsubscript{4} attains an agreement score of \( 0.576 \), while the Hawkes arrival ratio attains \( 0.413 \). In the highest-persistence bin (runs of at least 59 windows), the ordering reverses: the Hawkes arrival-ratio rule reaches \( 0.702 \) agreement with the oracle, while CLF\textsubscript{4} attains \( 0.641 \).

\begin{figure}[!htbp]
  \centering

  \begin{minipage}[t]{\textwidth}
  \centering
  \begin{tikzpicture}[x=2.0cm,y=2.50cm,>=latex,
                      every node/.style={font=\footnotesize}]
    \definecolor{hawkesCol}{RGB}{57,106,177} 
    \definecolor{clfCol}{RGB}{218,124,48}    

    \def\Hqone{0.4060}
    \def\Cqone{0.4586}
    \def\Hqtwo{0.4890}
    \def\Cqtwo{0.5237}
    \def\Hqthree{0.4131}
    \def\Cqthree{0.5758}
    \def\Hqfour{0.7021}
    \def\Cqfour{0.6406}

    \draw[->] (0,0) -- (0,0.85) node[above] {Agreement score};
    \draw[->] (0,0) -- (4.3,0);

    \foreach \y in {0.0,0.2,0.4,0.6,0.8} {
      \draw[densely dotted,gray!60] (0,\y) -- (4.3,\y);
      \node[left] at (0,\y) {\footnotesize \y};
    }

    \def\xQone{0.7}
    \def\xQtwo{1.7}
    \def\xQthree{2.7}
    \def\xQfour{3.7}
    \def\barW{0.30}

    \draw[fill=hawkesCol!70,draw=black]
      ({\xQone-\barW},0) rectangle ({\xQone-0.05},\Hqone);
    \draw[fill=clfCol!80,draw=black]
      ({\xQone+0.05},0) rectangle ({\xQone+\barW},\Cqone);

    \draw[fill=hawkesCol!70,draw=black]
      ({\xQtwo-\barW},0) rectangle ({\xQtwo-0.05},\Hqtwo);
    \draw[fill=clfCol!80,draw=black]
      ({\xQtwo+0.05},0) rectangle ({\xQtwo+\barW},\Cqtwo);

    \draw[fill=hawkesCol!70,draw=black]
      ({\xQthree-\barW},0) rectangle ({\xQthree-0.05},\Hqthree);
    \draw[fill=clfCol!80,draw=black]
      ({\xQthree+0.05},0) rectangle ({\xQthree+\barW},\Cqthree);

    \draw[fill=hawkesCol!70,draw=black]
      ({\xQfour-\barW},0) rectangle ({\xQfour-0.05},\Hqfour);
    \draw[fill=clfCol!80,draw=black]
      ({\xQfour+0.05},0) rectangle ({\xQfour+\barW},\Cqfour);

    \node[above] at (\xQone-\barW+0.15,\Hqone) {\tiny 0.4060};
    \node[above] at (\xQone+\barW-0.15,\Cqone) {\tiny 0.4586};

    \node[above] at (\xQtwo-\barW+0.15,\Hqtwo) {\tiny 0.4890};
    \node[above] at (\xQtwo+\barW-0.15,\Cqtwo) {\tiny 0.5237};

    \node[above] at (\xQthree-\barW+0.15,\Hqthree) {\tiny 0.4131};
    \node[above] at (\xQthree+\barW-0.15,\Cqthree) {\tiny 0.5758};

    \node[above] at (\xQfour-\barW+0.15,\Hqfour) {\tiny 0.7021};
    \node[above] at (\xQfour+\barW-0.15,\Cqfour) {\tiny 0.6406};

    \node[below=8pt,align=center] at (\xQone,0)
      {$Q_1$\\(1--3 windows,\\ $n=133$)};
    \node[below=8pt,align=center] at (\xQtwo,0)
      {$Q_2$\\(4--10 windows,\\ $n=317$)};
    \node[below=8pt,align=center] at (\xQthree,0)
      {$Q_3$\\(11--58 windows,\\ $n=1{,}622$)};
    \node[below=8pt,align=center] at (\xQfour,0)
      {$Q_4$\\($\geq 59$ windows,\\ $n=32{,}884$)};

    \draw[fill=hawkesCol!70,draw=black] (0.2,0.73) rectangle (0.4,0.81);
    \node[right] at (0.45,0.77) {Hawkes arrival ratio};

    \draw[fill=clfCol!80,draw=black] (0.2,0.62) rectangle (0.4,0.70);
    \node[right] at (0.45,0.66) {CLF\textsubscript{4}};
  \end{tikzpicture}

  \vspace{4pt}
  {\footnotesize (a) Agreement by benchmark-persistence bin}
\end{minipage}

  \vspace{10pt}

  \begin{minipage}[t]{\textwidth}
    \centering
    \begin{tikzpicture}[x=1.85cm,y=4.50cm,>=latex,
                        every node/.style={font=\footnotesize}]
      \definecolor{posCol}{RGB}{0,153,0}   
      \definecolor{negCol}{RGB}{204,76,76} 

      \def\Dqone{-0.0526}
      \def\Dqtwo{-0.0347}
      \def\Dqthree{-0.1628}
      \def\Dqfour{0.0615}

      \draw[->] (0,-0.22) -- (0,0.08) node[above] {$\text{Hawkes} - \text{CLF\textsubscript{4}}$};
      \draw[->] (0,0) -- (4.5,0);

      \foreach \y in {-0.20,-0.10,0.0,0.05} {
        \draw[densely dotted,gray!60] (0,\y) -- (4.5,\y);
        \node[left] at (0,\y) {\footnotesize \y};
      }

      \def\xQone{0.75}
      \def\xQtwo{1.8}
      \def\xQthree{2.9}
      \def\xQfour{4.0}
      \def\barW{0.30}

      \draw[fill=negCol!80,draw=black]
        ({\xQone-\barW},0) rectangle ({\xQone+\barW},\Dqone);
      \draw[fill=negCol!80,draw=black]
        ({\xQtwo-\barW},0) rectangle ({\xQtwo+\barW},\Dqtwo);
      \draw[fill=negCol!80,draw=black]
        ({\xQthree-\barW},0) rectangle ({\xQthree+\barW},\Dqthree);
      \draw[fill=posCol!70,draw=black]
        ({\xQfour-\barW},0) rectangle ({\xQfour+\barW},\Dqfour);

      \node[below=11pt,align=center] at (\xQone,0) {$Q_1$};
      \node[below=10pt,align=center] at (\xQtwo,0) {$Q_2$};
      \node[below=10pt,align=center] at (\xQthree,0) {$Q_3$};
      \node[below=10pt,align=center] at (\xQfour,0) {$Q_4$};

      \node[below=13pt,align=center] at (\xQone,\Dqone) {\scriptsize $-0.0526$};
      \node[below=15pt,align=center] at (\xQtwo,\Dqtwo) {\scriptsize $-0.0347$};
      \node[below=1pt,align=center] at (\xQthree,\Dqthree) {\scriptsize $-0.1628$};
      \node[above=4pt,align=center] at (\xQfour,\Dqfour) {\scriptsize $+0.0615$};

    \end{tikzpicture}

    \vspace{4pt}
    {\footnotesize (b) Difference in agreement (Hawkes $-$ CLF\textsubscript{4}) by bin}
  \end{minipage}

  \caption{Agreement scores of the Hawkes arrival-ratio and CLF\textsubscript{4} across benchmark-persistence bins. Panel~(a) shows agreement relative to the oracle in each bin of benchmark persistence. Panel~(b) displays the corresponding difference in agreement (Hawkes minus CLF\textsubscript{4}); CLF\textsubscript{4} attains higher agreement in the three lower-persistence bins (negative bars), while the Hawkes arrival ratio attains higher agreement in the highest-persistence bin (positive bar).}
  \label{fig:benchmark_persistence_accuracy}
\end{figure}

The aggregate agreement scores, the benchmark-persistence and contract-wise decompositions describe a consistent pattern. The Hawkes arrival-ratio rule and the LOB-based CLF\textsubscript{4} rule both achieve non-trivial agreement with the market-optimal oracle, with a large intersection of windows in which they select the same reference leg, yet they extract information from different parts of the joint distribution of order-flow and LOB state. The persistence analysis shows that most evaluation windows fall into long benchmark regimes, within which the arrival-ratio rule is comparatively more accurate, while the CLF\textsubscript{4} rule performs relatively better in short and medium-lived regimes in which the oracle switches reference legs more frequently. The contract-wise panels further indicate that the arrival ratio is most informative when the benchmark favours the current-month contract \(F_c\), whereas CLF\textsubscript{4} contributes more when the benchmark favours the next-month contract \(F_n\). These findings suggest that event-history and LOB-state information are complementary rather than interchangeable inputs for reference-leg selection. Figures~\ref{fig:benchmark_persistence_runs} and~\ref{fig:hawkes_clf4_joint} summarise this structure, including the contract-wise split of runs across the two contracts. They motivate the state-dependent interpretation of the two complementary signals developed in the discussion that follows.

\section{Discussion}
\label{sec:discussion}

The prevailing literature on market microstructure supports such \emph{state-dependent} performance. Empirical studies show that LOB shape, depth profiles and queue imbalance near the best quotes carry information about short horizon price impact  \cite{bouchaud2002statistical,cont2010price}. The CLF is computed directly from limit order book primitives: it responds to instantaneous depth, queue imbalance, and local LOB shape. This design allows it to adjust quickly when the relative liquidity of the two legs changes over short horizons. At the same time, during periods of intense trading activity, high-frequency quotes and trades are subject to microstructure noise and fleeting orders. Hasbrouck \cite{hasbrouck2013low} emphasises that low-latency trading can amplify such transitory effects in the observed LOB. Under such conditions, a purely cross-sectional LOB-based statistic may produce spurious or short-lived signals.

The Hawkes arrival-ratio, in contrast, is a low-dimensional summary of the event-time dynamics that influence price-impacting events. It aggregates information about how often price-impacting events arrive on each leg, with an explicit memory structure. Under standard Hawkes specifications, intensity forecasts are driven by self- and cross-excitation in recent order-flow \cite{bacry2015hawkes}. This construction makes the arrival ratio relatively insensitive to very short-lived fluctuations in the order-flow, and better suited to detecting and exploiting \emph{persistent} directional regimes in the benchmark.

The pattern in benchmark persistence aligns with this microstructure interpretation. In segments with low benchmark persistence, where the oracle switches legs frequently or after only short bursts, CLF\textsubscript{4} achieves higher agreement with the oracle than the Hawkes arrival ratio. In the highest-persistence bin, where the benchmark exhibits long, stable regimes, the Hawkes arrival ratio attains the higher agreement score. Collapsing the bins into low-persistence (runs of length at most 58 windows) and high-persistence (runs of at least 59 windows) segments reproduces the aggregate pattern discussed above: in the low-persistence segment CLF\textsubscript{4} attains an agreement score of \(0.5155\) compared with \(0.3903\) for the Hawkes arrival ratio, whereas in the high-persistence segment the Hawkes arrival ratio reaches \(0.7060\) compared with \(0.6441\) for CLF\textsubscript{4}. In this sense, the two rules exploit complementary parts of the benchmark’s temporal structure: short-memory cross-sectional signals versus longer-memory event-history signals.

\section{Conclusion}
\label{sec:conclusion}

This study develops and evaluates a diagnostic framework for reference-leg selection in multi-contract quoting. Using tick-by-tick NIFTY index futures data, we construct a market-optimal oracle based on realised slippage and compare two reference-leg decision rules: a Hawkes arrival-ratio rule built from reference-impacting order-flow, and a CLF-based rule constructed from contemporaneous LOB shape. In our illustrative application, the Hawkes arrival ratio attains higher agreement with the oracle in high-persistence benchmark regimes, while the CLF-based rule is relatively more accurate when the benchmark switches frequently; in addition, there is a non-trivial set of windows in which both rules match the oracle, indicating that event-history and LOB-state information often point to the same market-optimal benchmark.

Methodologically, the framework is diagnostic in scope. It is parsimonious at runtime and portable across tick-driven, multi-contract markets where quoting is employed,since it relies only on tick-time event histories and LOB snapshots.The diagnostic emphasis is deliberate: it isolates decision quality from strategy-specific implementation details and clarifies where event-history and LOB-state information each contribute to reference-leg selection. Evaluation is anchored by an infeasible hindsight lower bound on deviation from the target spread, which provides a common yardstick for comparing heterogeneous rules, and all tests follow a walk-forward, out-of-sample protocol that relies on model-comparison procedures controlling for data-snooping risk \cite{hansen2005test,2000a}.

The construction of the CLF places the framework within the strand of microstructure research that links LOB state to short-horizon execution risk. The incorporation of cross-level log-quote slopes is motivated by Hasbrouck’s analysis of low-latency order books, where steep slopes in the best quotes are associated with fragile depth and heightened sensitivity to incoming orders \cite{hasbrouck2013low}. In this paper these ingredients are aggregated into a single, direction-sensitive score per leg, designed to be computed from the LOB without repeated high-dimensional re-estimation.

The Hawkes arrival ratio connects the study to intensity-based models of execution risk with limit orders. The prevailing literature investigates limit-order fills and price moves modeled by point-process intensities~\cite{gueant2013inventory,cartea2015algorithmic,guilbaud2013limitmarket}. The associated execution and price risk is managed by steering these intensities rather than by reacting solely to static book configurations. However, these models are constructed for single contract order-flow and do not naturally extend to multi-contract quoting frameworks and the dependent nature of execution risk in such a setting. The arrival ratio used here is a reduced-form summary of such intensity forecasts, the framework uses their relative magnitudes to choose the leg that is less exposed to future price-impacting flow. In our illustrative application, the non-trivial intersection of windows in which both CLF and Hawkes arrival ratios match the oracle suggests that cross-sectional book shape and short-run order-flow clustering often point to the same market-optimal oracle.

Looking ahead, the same diagnostic structure can be embedded in an explicit control formulation. In such a setting, each quoting rule induces a mapping from an event-time state (summarising recent order-flow, LOB shapes or other market co-variates) to a reference-contract choice. The oracle benchmark provides a natural loss functional based on slippage relative to the choice of market optimal reference contract. Formalising this extension is beyond the scope of the present methodological study, but the results here suggest that combining event-history and LOB-state signals in such a control framework is a promising direction for future work.

\bibliography{references}

\include{appendix_1}
\include{appendix_2}
\include{appendix_3}

\end{document}

%% file: appendix_1.tex
\section{Appendix I: Hawkes excitation kernel norms}

We visualize the Hawkes EM kernel norms across three time blocks to illustrate how excitation strength between event types evolves intraday. Each pair shows the ask and bid sides during a 20-minute window.

\begin{figure}[!htbp] 
  \centering
  \captionsetup{justification=raggedright,singlelinecheck=false}
  \setlength{\abovecaptionskip}{6pt}
  \setlength{\belowcaptionskip}{4pt}
  \setlength{\textfloatsep}{8pt}
  \setlength{\floatsep}{8pt}

  \begin{subfigure}[t]{0.48\textwidth}
    \centering
    \includegraphics[width=\linewidth]{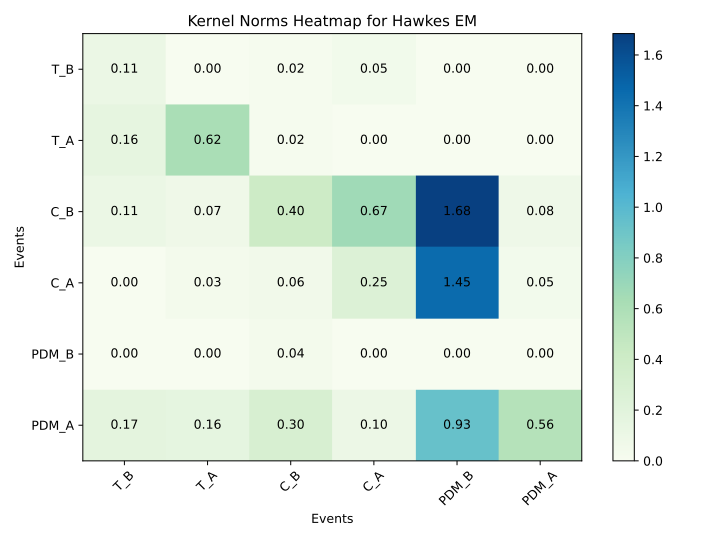}
    \caption{\textit{Fig. A1. 10:05--10:25 AM: Strong self-excitation in price-down modifications (PDM), NIFTY Feb.}}
  \end{subfigure}\hfill
  \begin{subfigure}[t]{0.48\textwidth}
    \centering
    \includegraphics[width=\linewidth]{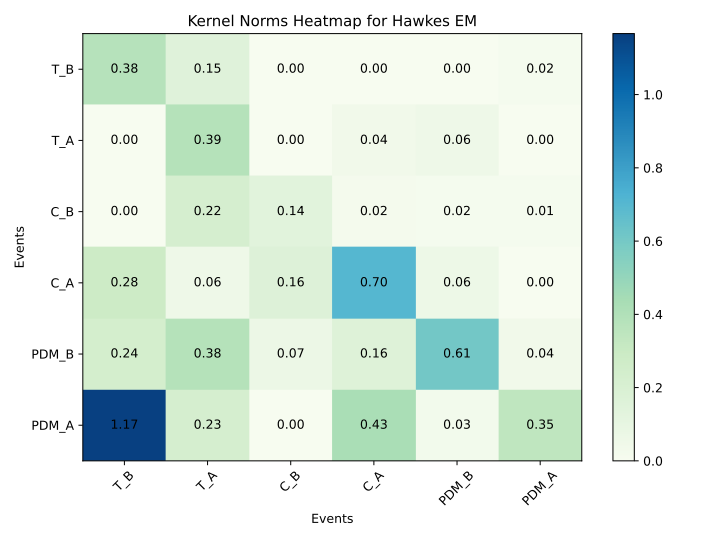}
    \caption{\textit{Fig. A2. 10:05--10:25 AM: Strong self-excitation in price-down modifications (PDM), NIFTY Mar.}}
  \end{subfigure}

  \vspace{0.6em}

  \begin{subfigure}[t]{0.48\textwidth}
    \centering
    \includegraphics[width=\linewidth]{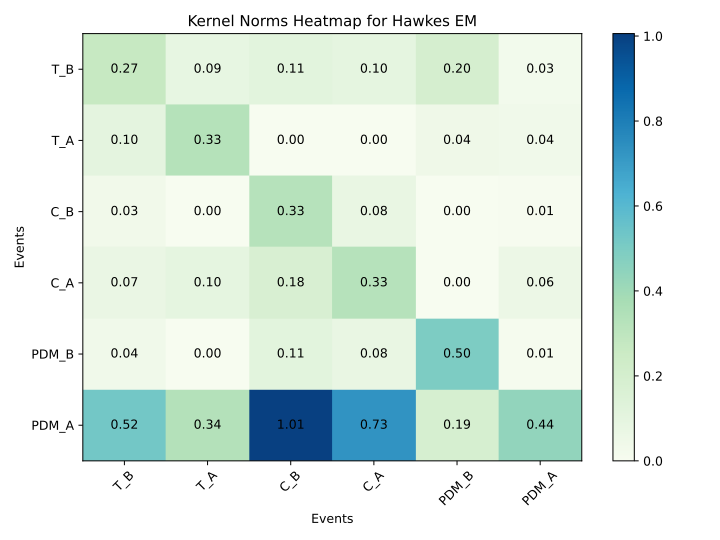}
    \caption{\textit{Fig. A3. 12:10--12:30 PM: Increased excitation from trades and cancellations, NIFTY Feb.}}
  \end{subfigure}\hfill
  \begin{subfigure}[t]{0.48\textwidth}
    \centering
    \includegraphics[width=\linewidth]{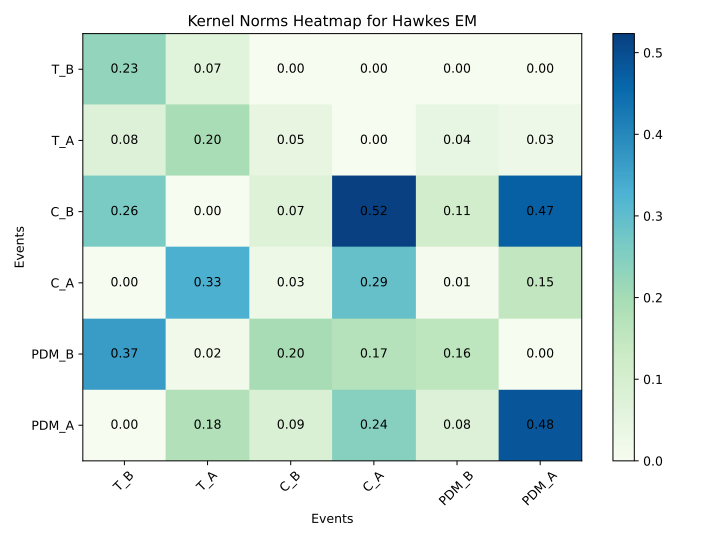}
    \caption{\textit{Fig. A4. 12:10--12:30 PM: Increased excitation from trades and cancellations, NIFTY Mar.}}
  \end{subfigure}

  \vspace{0.6em}

  \begin{subfigure}[t]{0.48\textwidth}
    \centering
    \includegraphics[width=\linewidth]{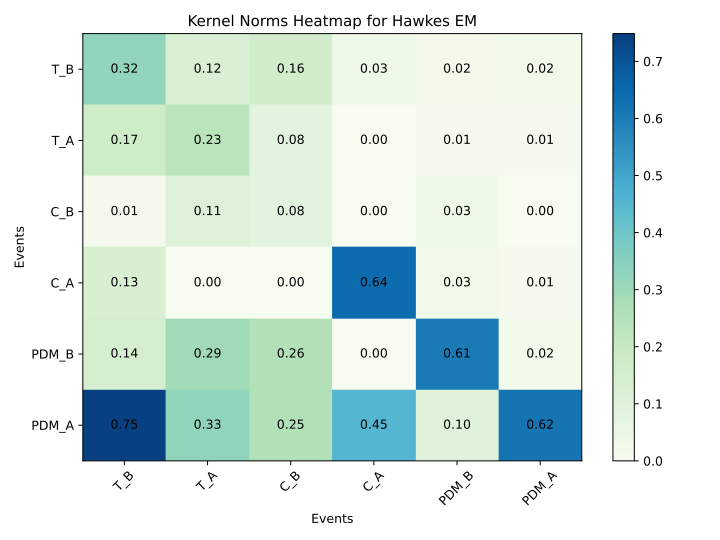}
    \caption{\textit{Fig. A5. 2:10--2:30 PM: Cancellations dominate excitation; PDM influence fades, NIFTY Feb.}}
  \end{subfigure}\hfill
  \begin{subfigure}[t]{0.48\textwidth}
    \centering
    \includegraphics[width=\linewidth]{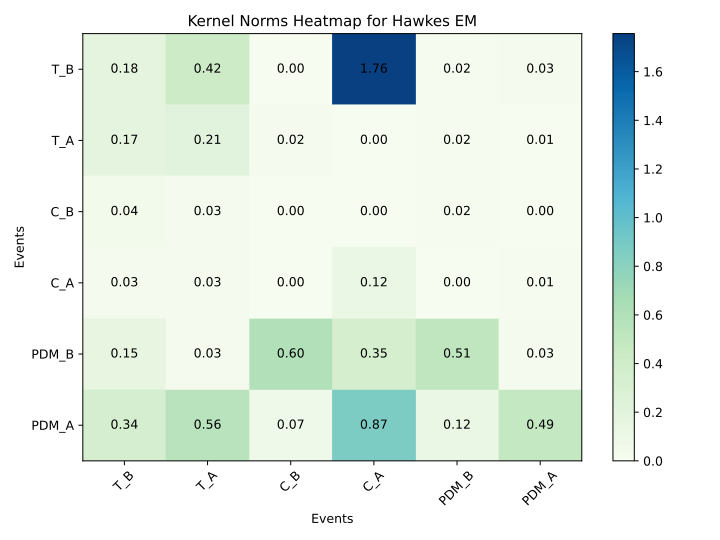}
    \caption{\textit{Fig. A6. 2:10--2:30 PM: Cancellations dominate excitation; PDM influence fades, NIFTY Mar.}}
  \end{subfigure}

  \caption{Excitation diagnostics by interval and contract. Panels are arranged two per row and scaled to fit on a single float page.}
  \label{fig:excitation-grid}
\end{figure}

\hfill \break
\hfill \break
In each pair of kernel norm heat maps shown, the first heat map is for the NIFTY futures contract expiring on 24th February 2022, while the second heatmap is for the NIFTY futures contract expiring on March 30th 2022. 

\begin{table}[!ht]
\renewcommand\thetable{A1}
\centering
\caption{Microstructure Event Types Used in Hawkes Modeling}
\label{tab:event_types}
\begin{tabular}{ll}
\toprule
\textbf{Label} & \textbf{Description} \\
\midrule
$T_A$   & Trade from aggressive order on the ask. \\
$T_B$   & Trade from aggressive order on the bid. \\
$C_A$   & Cancellation at the top level of the ask. \\
$C_B$   & Cancellation at the top level of the bid. \\
$PDM_A$ & Downward price adjustment on the ask. \\
$PDM_B$ & Downward price adjustment on the bid. \\
\bottomrule
\end{tabular}
\end{table}

%% file: appendix_2.tex
\section{Appendix II: Sample Tick Data}

The table below displays a snippet of the raw tick-by-tick data used in the experiments. It includes timestamps, event types, order sides, price levels, and available quantities. This example illustrates the granularity and structure of the input used to estimate Hawkes kernels and compute liquidity metrics.

\clearpage
\justifying

\begin{sidewaystable}[!htbp]
    \centering
    \renewcommand\thetable{A2}
    \begingroup
    \footnotesize
    \setlength{\tabcolsep}{3pt}
    \renewcommand{\_}{\textunderscore\discretionary{}{}{}}
    \begin{tabular}{
        p{2.8cm}  
        p{2.2cm}  
        p{1.6cm}  
        p{0.6cm}  
        p{1.0cm}  
        p{0.9cm}  
        p{2.2cm}  
        p{2.0cm}  
        p{1.5cm}  
        p{3.2cm}  
    }
        \toprule
        \textbf{Server\_epoch\_nanos} & \textbf{Symbol} & \textbf{event\_type} & \textbf{side} & \textbf{Price} & \textbf{Qty} & \textbf{OID1} & \textbf{OID2} & \textbf{ Aggressor} & \textbf{capture\_timestampz} \\
        \midrule
        1644227998953429992 & NSEFNO\_NIFTY\_G22 & TRADE &  & 17214.00 & 50  & 1100000071295344 & 1100000071290667 & \    BUY & 2022-02-07 09:59:58.953429992+00 \\
        1644227999031664383 & NSEFNO\_NIFTY\_G22 & TRADE &  & 17213.80 & 50  & 1100000071295449 & 1100000071295363 & \ BUY & 2022-02-07 09:59:59.031664383+00 \\
        1644227999031664383 & NSEFNO\_NIFTY\_G22 & TRADE &  & 17213.85 & 150 & 1100000071295449 & 1100000071290493 & \ BUY & 2022-02-07 09:59:59.031664383+00 \\
        \midrule
        1644205500049497674 & NSEFNO\_NIFTY\_G22 & N & BUY  & 17000.00 & 50  & 1100000000001754 &  &  & 2022-02-07 03:45:00.049497674+00 \\
        1644205500050566579 & NSEFNO\_NIFTY\_G22 & N & BUY  & 17401.10 & 100 & 1100000000001939 &  &  & 2022-02-07 03:45:00.050566579+00 \\
        1644205500050579855 & NSEFNO\_NIFTY\_G22 & N & BUY  & 16901.05 & 100 & 1100000000001941 &  &  & 2022-02-07 03:45:00.050579855+00 \\
        \midrule
        1644205500050016996 & NSEFNO\_NIFTY\_G22 & N & SELL & 17460.00 & 50  & 1100000000001849 &  &  & 2022-02-07 03:45:00.050016996+00 \\
        1644205500050096128 & NSEFNO\_NIFTY\_G22 & N & SELL & 17500.00 & 100 & 1100000000001864 &  &  & 2022-02-07 03:45:00.050096128+00 \\
        1644205500050121328 & NSEFNO\_NIFTY\_G22 & N & SELL & 17458.55 & 400 & 1100000000001868 &  &  & 2022-02-07 03:45:00.050121328+00 \\
        \midrule
        1644227999966936439 & NSEFNO\_NIFTY\_G22 & MODIFY\_TICK & SELL & 17225.00 & 250 & 1100000045070355 &  &  & 2022-02-07 09:59:59.966936439+00 \\
        1644227999969919000 & NSEFNO\_NIFTY\_G22 & MODIFY\_TICK & BUY  & 17205.95 & 100 & 1100000071271630 &  &  & 2022-02-07 09:59:59.969919000+00 \\
        1644227999988916651 & NSEFNO\_NIFTY\_G22 & MODIFY\_TICK & SELL & 17216.70 & 100 & 1100000071286370 &  &  & 2022-02-07 09:59:59.988916651+00 \\
        \midrule
        1644228205338970644 & NSEFNO\_NIFTY\_G22 & CANCEL\_TICK & SELL & 17444.00 & 50  & 1100000012222592 &  &  & 2022-02-07 10:03:25.338970644+00 \\
        1644228205340990944 & NSEFNO\_NIFTY\_G22 & CANCEL\_TICK & SELL & 17273.00 & 50  & 1100000010068455 &  &  & 2022-02-07 10:03:25.340990944+00 \\
        1644228205341822198 & NSEFNO\_NIFTY\_G22 & CANCEL\_TICK & BUY  & 17000.00 & 550 & 1100000058552827 &  &  & 2022-02-07 10:03:25.341822198+00 \\
        \bottomrule
    \end{tabular}
    \endgroup
    \caption{Data snippet for 2022-02-07 (\texttt{TRADE}: trade tick order, \texttt{N}: new tick order, \texttt{MODIFY\_TICK}: modification tick order, \texttt{CANCEL\_TICK}: cancellation tick order).}
    \label{tab:trade_data}
\end{sidewaystable}

%% file: appendix_3.tex
\section{Appendix III: Global Notation}
\begin{table}[!htbp]
\centering
\caption{Summary of Symbols and Their First Point of Introduction}
\label{tab:global_notation_clean}
\begin{tabular}{p{3cm}p{7cm}p{3cm}}
\toprule
\textbf{Symbol} & \textbf{Description} & \textbf{Section} \\
\midrule
\multicolumn{3}{l}{\textit{Market Structure and Events}} \\
$F_c$, $F_n$ & Current and next-month futures contracts & Problem Formulation \\
$p^{b,i}, p^{a,i}$ & Price at level $i$ on bid / ask side & CLF Section \\
$q^{b,i}, q^{a,i}$ & Quantity at level $i$ on bid / ask side & CLF Section \\
$X \in \{c, n\}$ & Contract index: current or next-month & Forecasting Objective \\
$Y \in \{b, a\}$ & Limit order book side: bid or ask & Forecasting Objective \\
$\mathcal{E}$ & Event types: trades (T), cancels (C), PDM & Forecasting Objective \\
$T_A, T_B$ & Trades due to aggressive orders on ask/bid & Appendix Table \\
$C_A, C_B$ & Cancellations at top of ask/bid & Appendix Table \\
$PDM_A, PDM_B$ & Price-down moves on ask/bid side & Appendix Table \\
\midrule

\multicolumn{3}{l}{\textit{Feature Extraction and Data Structures}} \\
$\mathcal{T}(w)$ & Tick history in interval $w$ & Problem Formulation \\
$\mathcal{B}(w)$ & LOB snapshot at start of window $w$ & Problem Formulation \\
$\mathcal{W}$ & Set of all evaluation windows $w = [\tau, \tau + \xi)$ & Problem Formulation \\
$\tau$ & Start time of evaluation window $w$ & Problem Formulation \\
$\xi$ & Forecast horizon length & Problem Formulation \\
$\mathcal{G}^{\mathcal{T}}, \mathcal{G}^{\mathcal{B}}, \mathcal{G}^{\mathcal{H}}$ & Resulting feature vectors from each input source & Problem Formulation \\
\midrule

\multicolumn{3}{l}{\textit{Decision Rules and Outcomes}} \\
$\chi^{\mathcal{T}}(w), \chi^{\mathcal{B}}(w), \chi^{\mathcal{H}}(w)$ & Binary decisions: 0 = $F_c$, 1 = $F_n$ & Problem Formulation \\
$\chi^m(w)$ & Market-optimal reference leg (minimum realized spread) & Problem Formulation \\
\midrule

\multicolumn{3}{l}{\textit{Hawkes Process Components}} \\
$\lambda^i_t$ & Intensity function for event type $i$ & Hawkes Forecasting \\
$\mu^i$ & Baseline intensity for event type $i$ & Hawkes Forecasting \\
$\phi_{ij}(t)$ & Kernel function: excitation from $j$ to $i$ & Hawkes Forecasting \\
$\tilde{N}^{X,Y}_{(w)}$ & Simulated total event count over window $w$ & Forecasting Objective \\
$\tilde{N}^{X,Y,r}_{(w)}$ & Simulated reference-impacting event count & Forecasting Objective \\
$\rho^{X,Y}_{(w)}$ & Arrival ratio: $\tilde{N}^{X,Y,r}_{(w)} / \tilde{N}^{X,Y}_{(w)}$ & Forecasting Objective \\
\midrule

\multicolumn{3}{l}{\textit{Liquidity and Evaluation Metrics}} \\
$CLF_i$ & Composite Liquidity Factor using top $i$ LOB levels & CLF Section \\
$\delta_S^{(r)}(w)$ & Realized spread under contract $r$ as reference leg & Problem Formulation \\
$\mathcal{A}(\chi, \chi^m)$ & Agreement score: $\frac{1}{|\mathcal{W}|} \sum \mathbb{I}\{\chi(w) = \chi^m(w)\}$ & Problem Formulation \\
\bottomrule
\end{tabular}
\end{table}